\documentclass{emulateapj}
\usepackage{apjfonts}

\begin{document}

\newcommand{\vdag}{(v)^\dagger}
\newcommand{\oii}{[O{\sc ii}] }
\newcommand{\sigss}{${\Sigma}_{2D}$}
\newcommand{\sigsss}{${\Sigma}_{3D}$}
\newcommand{\pog}{Poggianti et al. (2006) }




\shorttitle{Star formation-local density relation}
\shortauthors{Poggianti et al.}




\title{The relation between star formation, morphology and local density 
       in high redshift clusters and groups\footnote{Based on observations collected at the European Sputhern Observatory, Chile, as part of large programme 166.A-0162 (the ESO Distant Cluster Survey) and with the NASA/ESA Hubble Space Telescope with proposal 9476.}
}

\author{Bianca M. Poggianti,
Vandana Desai, Rose
Finn, Steven Bamford, Gabriella De Lucia, Jesus Varela,
Alfonso Aragon-Salamanca, Claire Halliday, Stefan Noll,
Roberto Saglia, Dennis Zaritsky, Philip Best,
Douglas Clowe, Bo Milvang-Jensen, Pascale
Jablonka, Roser Pello, Gregory Rudnick, Luc
Simard, Anja von der Linden, Simon White}
\email{bianca.poggianti@oapd.inaf.it}




\begin{abstract}
We investigate how the [OII] properties and the morphologies of
galaxies in clusters and groups at $z=0.4-0.8$ depend on projected
local galaxy density.  We also compare our results with those derived
from the field at similar redshifts and from clusters at low-z. In
both nearby and distant clusters, higher-density regions contain
proportionally fewer star-forming galaxies.  At odds with low-z
results, the relation between star-forming fraction and local density
at high-z varies from high- to low-mass
clusters, although is not solely a function of cluster mass.
Both at high- and low-z, the average [OII] equivalent width (EW) of
star-forming galaxies is 
independent of local
density.  However, in distant clusters the average current star
formation rate in star-forming galaxies seems to
peak at densities $\sim 15-40$ galaxies Mpc$^{-2}$.
When using galaxy samples with similar mass distributions, we do not
find large variations in the average EW(OII) and SFR properties of
star-forming galaxies in the distant field, poor groups, and clusters.
Overall, these results suggest that at high-z the current star
formation activity in star-forming galaxies of a given galaxy mass
does not depend strongly on global or local environment, though
the possible SFR peak at intermediate densities seems at odds with
this conclusion.  
We derive cluster-integrated star
formation rates and find 
that the cluster SFR normalized by cluster mass anticorrelates with
mass and correlates with the star-forming fraction, although we
caution that the anticorrelation with mass could be mainly driven by
correlated errors.  We demonstrate that these trends can be understood
given a) that the average star-forming galaxy forms about one solar
mass per year (uncorrected for dust) in all our clusters;
b) how the total number of galaxies
scales with cluster mass; and c) the dependence of star-forming
fraction on cluster mass.  These findings show that the most
fundamental open question currently remains why the evolution of the
star-forming fraction depends on cluster mass.
We present the morphology-density relation for our
$z=0.4-0.8$ clusters, and uncover that the decline of the spiral fraction
with density is entirely driven by galaxies of types Sc or later,
while early spirals (Sa's and Sb's) have a flat distribution with
density, similarly to that of S0s.  For galaxies of a given Hubble
type, we see no evidence that star formation properties (ongoing SFR,
[OII] equivalent width, 4000 \AA $\,$ break) depend on local
environment.  In contrast with recent findings at low-z, in our
distant clusters the star formation-density relation and the
morphology-density relation are equivalent, suggesting that neither of
the two relations is more fundamental than the other.  We suggest that
the contrasting results at high- and low-z are partly due to different
classification methods and/or to evolutionary effects giving rise to a
progressive decoupling of the two relations at recent epochs.
\end{abstract}





\section{Introduction}

The star formation activity and other fundamental galaxy properties
such as morphology and gas content
vary systematically with redshift, galaxy mass and environment. While
redshifts and galaxy masses are at least conceptually clearly defined,
the definition of a galaxy environment is arbitrary and its optimal
choice would require an a priori knowledge of the very thing we are
trying to identify, that is the physical environmental driver or
drivers of galaxy formation and evolution.  Most studies nowadays
define the ``environment'' either in terms of the local galaxy number
density (the number of galaxies per unit volume or projected area
around the galaxy of interest), or the virial mass of the cluster or
group to which the galaxy belongs, when this applies, because these
are the two most easily measurable quantities from spectroscopic or
even imaging surveys.

In spite of environment being an elusive and arbitrary concept, the
fact that galaxy properties depend on environment was recognized
earlier than the dependence on galaxy mass and redshift (Hubble \&
Humason 1931).  The first quantitative measurement of systematic
differences with local environment was the so-called
morphology-density relation (MDR) in nearby clusters.  The MDR is the
observed variation of the proportion of different Hubble types with
local density, with ellipticals being more common in high-density
regions, spirals being more common in low-density regions, and S0s
making up a constant fraction of the total population within the
cluster virial radius regardless of density (Dressler 1980, as
revisited in Dressler et al. 1997). It was subsequently found that a
similar MDR also exists in nearby groups (Postman \& Geller 1984 also
revisited in Postman et al. 2005) and that a qualitatively similar,
but quantitatively different, MDR is present in galaxy clusters at
redshifts up to 1 (Dressler et al. 1997, Treu et al. 2003, Smith et
al. 2005, Postman et al. 2005).

Galaxy stellar populations have also long been known to vary
systematically with environment (Spitzer \& Baade 1951): denser
environments have on average older stellar populations. At least at
some level, this must be related to the higher incidence of early-type
galaxies in high density regions, i.e. to the morphology-density
relation.  At low redshift the best characterization of the ``star
formation-local density'' (SFD) relation has come from large redshift
surveys.  These studies have conclusively demonstrated that the
average galaxy properties related to star formation depend on local
density even at large clustercentric radii; at low densities in
clusters; and outside of clusters, in groups and the general field
(Hashimoto et al. 1998, Lewis et al. 2002, Gomez et al. 2003,
Kauffmann et al. 2004, Balogh et al. 2004, see also Pimbblet et
al. 2002).  Moreover, they have highlighted the dependence on both
galaxy mass and local environment, showing strong environmental trends
at a given galaxy mass (Kauffmann et al. 2004, Baldry et al. 2006).

To understand the origin of the SFD relation, it is essential to
answer two separate questions: at any given galaxy mass, 1) how does
the proportion of star-forming galaxies vary with density, and 2) how
does the star formation activity in star-forming galaxies vary with
density?  At low-z, the evidence for a change in the relative numbers
of red/passively-evolving and blue/star-forming galaxies with local
environment is overwhelming, but it remains an open question whether
star-forming galaxies of similar mass have star formation histories
that depend on local density (Balogh et al. 2004a,b, Hogg et al. 2004,
Gomez et al. 2003, Baldry et al. 2006).

Deep redshift surveys have recently extended the study of the SFD
relation in the general field to high redshift.  They have revealed
that the number ratio of red to blue galaxies increases with local density
out to $z>1$ and that we might be witnessing the establishment of the
color-density relation at $z$ approaching 1.5 (Cucciati et al. 2006,
Cooper et al. 2007, Cassata et al. 2007).  This suggests that the
transition from a star-forming phase to a passive one occurs for a
large number of massive galaxies in groups at $z\sim2$ (Poggianti et
al.~2006). In apparent but not substantial contradiction, the average
SFR per galaxy at $z=1$ increases instead of decreasing with local
density.  Therefore the SFD relation is inverted with respect to the
local Universe (Elbaz et al. 2007, Cooper et al. 2008).

Thus, the MDR up to $z=1$ is well studied in clusters and has started
to be explored in the field (Capak et al. 2007), and the SFD relation
is now being investigated in the general field over a similar redshift
baseline. In clusters, Moran et al. (2005) have presented the EW(OII)s
of ellipticals and S0 galaxies as a function of local density for a
cluster at $z=0.4$.  However, a detailed study of the relation between
star-formation and local environment in distant clusters has not yet
been carried out.  As a consequence, a comparison of the MDR and the
SFD relation has not been possible to date in clusters at high
redshift.  This is due to the limited number of well-studied distant
clusters with homogeneous data.

Deep galaxy redshift surveys have recently made it possible to
characterize the global as well as the local environment of galaxies
and to study significant samples of groups at high-z (Wilman et
al. 2005, Gerke et al. 2005, Balogh et al. 2007, Gerke et al. 2007,
Finoguenov et al. 2007). Groups are typically identified as galaxy
associations with masses $< 10^{14} \, M_{\odot}$ corresponding to
velocity dispersions of $ \lesssim 400 \, \rm km \, s^{-1}$.  Even the
largest field surveys, however, include only very few distant systems
above this mass (Finoguenov et al. 2007, Gerke et al. 2007).  On the
other hand, until recently, distant cluster surveys have studied
primarily massive clusters. Only
the latest surveys of optically-selected samples target structures of
a wide range of masses, down to the group level (Hicks et al. 2008,
Gilbank et al. 2008, Milvang-Jensen et al. 2008, Halliday et
al. 2004).
Nowadays groups are therefore the meeting point of field and cluster
studies at high-z, and it has become possible to study the dependence
of the SFD relation on global environment (clusters, groups and the
field), approaching the question from both perspectives.

In this paper, we investigate the relation between star formation
activity, morphology and local galaxy density in $z=0.4-0.8$ clusters
and groups observed by the ESO Distant Cluster Survey (EDisCS).  The
EDisCS data set permits an internal comparison with galaxies in poor
groups and the field at the same redshifts in a homogeneous way. To
compare our results with clusters in the nearby Universe, we use a
cluster sample drawn from the Sloan Digital Sky Survey (SDSS).

After presenting the data set (\S2) and the definition of our cluster,
group and field samples (\S3), we outline our method for measuring the
local galaxy density at high-z (\S4) and describe the low-z cluster
sample we use as local comparison (\S5).  The average trends of the
fraction of star-forming galaxies and of the [OII] equivalent width
with local density in clusters (\S6.1) is compared with those found in
lower density environments in \S6.2 and to those in low-z clusters in
\S6.4.  
The dependence of the star-forming
fraction on global cluster properties is presented in \S6.3.  We then
analyze the behavior of the average and specific star formation rates
in \S7, summarizing the similarities and differences of the EW-density
and SFR-density relations in \S7.1. Cluster-integrated star formation
rates are derived in \S7.2 where we show their relationship with
cluster mass and other global cluster properties.  Galaxy morphologies
are discussed in \S8, where we present the morphology-density
relation, the star-forming properties of each Hubble type as a
function of local density, and the link between the MD and the SFD
relations. The latter is compared with results at low redshift in
\S8.1. Finally, we summarize our conclusions in \S9.

All equivalent widths and cluster velocity dispersions are given in
the rest frame. All quantities related to
star formation are given
uncorrected for dust. We use proper (not comoving) radii, areas and
volumes.
We assume a $\Lambda$CDM cosmology
with ($H_0$, ${\Omega}_m$, ${\Omega}_{\lambda}$) = (70,0.3,0.7).

\section{The dataset}

The ESO Distant Cluster Survey (hereafter, EDisCS) is a
multiwavelength survey of galaxies in 20 fields containing galaxy
clusters at $z=0.4-1$.

Candidate clusters were selected as surface brightness peaks in
smoothed images taken with a very wide optical filter ($\sim$
4500-7500 \AA) as part of the Las Campanas Distant Cluster Survey
(LCDCS; Gonzales et al.~2001). The 20 EDisCS fields were chosen among
the 30 highest surface brightness candidates in the LCDCS, after
confirmation of the presence of an apparent cluster and of a possible
red sequence with VLT 20 min exposures in two filters (White et
al. 2005).

For all 20 fields EDisCS has obtained
deep optical photometry with FORS2/VLT, near-IR photometry with
SOFI/NTT, multislit spectroscopy with FORS2/VLT, and MPG/ESO 2.2/WFI
wide field imaging in $VRI$. 
ACS/HST mosaic imaging in $F814W$ of 10 of the highest
redshift clusters has also been acquired (Desai et al. 2007). Other 
follow-up programmes 
include XMM-Newton X-Ray observations (Johnson et al. 2006), Spitzer
IRAC and MIPS imaging (Finn et al. in prep.), $\rm H\alpha$
narrow-band imaging (Finn et al. 2005), and additional optical imaging
and spectroscopy in 10 of the EDisCS fields targeting galaxies at
$z\sim5$ (Douglas et al. 2007).

An overview of the survey goals and strategy is given by White et
al. (2005), who also present the optical ground--based photometry.
This consists of $V$, $R$ and $I$ imaging for the 10 highest redshift
cluster candidates, aimed at providing a sample at $z \sim 0.8$
(hereafter the high-z sample) and $B$, $V$ and $I$ imaging for 10
intermediate--redshift candidates, aimed to provide a sample at $z
\sim 0.5$ (hereafter the mid-z sample). In practice, the redshift
distributions of the high-z and the mid-z samples partly overlap
(Milvang-Jensen et al. 2008).

Spectra of $>100$ galaxies per cluster field were obtained, with
typical exposure times of 4 hours for the high-z sample and 2 hrs for
the mid-z sample.  Spectroscopic targets were selected from
\textit{I}-band catalogs (Halliday et al.  2004).  At the redshifts of
our clusters, this corresponds to $\sim 5000 \pm500$ \AA $\,$ rest
frame.  Conservative rejection criteria based on photometric redshifts
(Pell{\'o} et al. in prep.)  were used in the selection of spectroscopic
targets to reject a significant fraction of non--members while
retaining a spectroscopic sample of cluster galaxies equivalent to a
purely \textit{I}-band selected one. {\it A posteriori}, we verified
that these criteria have excluded at most 1-3\% of cluster galaxies
(Halliday et al. 2004 and Milvang-Jensen et al. 2008).  The
spectroscopic selection, observations, and catalogs are presented in
Halliday et al. (2004) and Milvang-Jensen et al. (2008).

In this paper we make use of the spectroscopic completeness weights
derived by Poggianti et al. (2006). Here we only give a brief summary
of the completeness of our spectroscopic sample, referring the reader
to the previous paper for details.  Given the long exposure times, the
success rate of our spectroscopy (number of redshift/number of spectra
taken) is 97\% above the magnitude limit used in this study. A visual
inspection of the remaining 3\% of the galaxies reveals that most of
these are bright, featureless low-z galaxies.  Moreover, in our
previous paper we computed the spectroscopic completeness as a
function of galaxy magnitude and position within the cluster (Appendix
A), verified the absence of biases in the completeness-corrected
sample (Appendix B) and found that incompleteness has a negligible
effect on the [OII] properties of our clusters.

In this paper, we analyze 16 of the 20 fields that comprise the
original EDisCS sample. We exclude two fields that lack several masks
of deep spectroscopy (cl1122.9-1136 and cl1238.5.114; see Halliday et
al. 2004 and White et al. 2005). We also exclude two additional
systems (cl1037.9-1243 and cl1103.7-1245), each of which has a
neighboring rich structure at a slightly different redshift
(Milvang-Jensen et al. 2008) that is indistinguishable on the basis of
photometric properties alone.  The names, redshifts, velocity
dispersions, and numbers of spectroscopic members for the remaining 16
clusters are listed in Table~1.

The EDisCS spectra have a dispersion of 1.32 \AA/pixel or 1.66
\AA/pixel, depending on the observing run.  They have a FWHM
resolution of $\sim 6$ \AA, corresponding to rest-frame 3.3 \AA $\,$
at z=0.8 and 4.3 \AA $\,$ at z=0.4.  The equivalent widths of \oii
were measured from the spectra using a line-fitting technique, as
outlined in Poggianti et al. (2006).  This method includes visual
inspection of each 1D spectrum.  Each line detected in a given 1D
spectrum was confirmed by visual inspection of the corresponding 2D
spectrum; this is especially useful to assess the reality of weak \oii
lines.

We do not attempt to separate a possible AGN contribution to the [OII]
line, or to exclude galaxies hosting an AGN.  We are unable to
identify AGNs in our data, as the traditional optical diagnostics are
based on emission lines that are not included in the spectral range
covered by most of our spectra.

AGN contamination will be most relevant for our study if there are
non-starforming, red galaxies in which the [OII] emission originates
exclusively from processes other than star formation.  We note that
only 13\% of the spectroscopic sample used here is composed of red
galaxies\footnote{We define as red a galaxy with a 4000 break $\geq
1.5$, see \S8.} with a detected [OII] line in emission. About 40\% of
these are spirals of types Sa or later.
Among local field galaxies, about half of the red [OII]-emitting
galaxies are LINERs whose source of ionization is still debated, while
the rest are either star-forming or Seyferts/transition objects in
which star formation dominates the line emission (Yan et al. 2006).
Adopting the distribution of AGN types observed locally in the field
and conservatively assuming that all LINERs are devoid of star
formation, we estimate that the contamination from pure AGNs in our
sample is at most 7\%. Moreover, we have verified that the fraction of
red emission-line galaxies is not a function of local density.  It
remains true, however, that all the trends we observe, and their
evolution, may reflect a combination of the variations in the level of
both star formation activity {\it and} AGN activity. This should be
kept in mind throughout the paper and when comparing our results with
any other work.

With these caveats in mind, we conveniently refer to galaxies
interchangeably as ``star-forming'' or ``\oii galaxies'' whenever
their EW(\oii)$>3$ \AA $\,$, adopting the convention that EWs are
positive in emission. The detection of the [OII] line above this EW
limit is essentially complete in our spectroscopic sample (see
Poggianti et al. 2006 for details).

\begin{table}
\begin{center}
{\scriptsize
\caption{List of clusters.\label{tbl1}}
\begin{tabular}{llclr}
\tableline\tableline
&&&& \\
Cluster name & Short name & $z$ & $\sigma$ $\pm{\delta}_{\sigma}$ & $N_{mem}$ \\
&&& $\rm km \, s^{-1}$ & \\
\tableline
 Cl\,1232.5-1250     &  Cl\,1232     & 0.5414  & 1080 $_{-89}^{+119}$ &54\\ 
 Cl\,1216.8-1201     &  Cl\,1216     & 0.7943  & 1018 $_{-77}^{+73}$  &67\\ 
 Cl\,1138.2-1133     &  Cl\,1138     & 0.4798  &  732 $_{-76}^{+72}$  &49\\ 
 Cl\,1411.1-1148     &  Cl\,1411     & 0.5201  &  710 $_{-133}^{+125}$&22\\ 
 Cl\,1301.7-1139     &  Cl\,1301     & 0.4828  &  687 $_{-86}^{+81}$  &35\\ 
 Cl\,1353.0-1137     &  Cl\,1353     & 0.5883  &  666 $_{-139}^{+136}$ &20\\ 
 Cl\,1354.2-1230     &  Cl\,1354     & 0.7627  &  648 $_{-110}^{+105}$ &21\\ 
 Cl\,1054.4-1146     &  Cl\,1054-11  & 0.6972  &  589 $_{-70}^{+78}$  &49\\ 
 Cl\,1227.9-1138     &  Cl\,1227     & 0.6355  &  574 $_{-75}^{+72}$  &22\\ 
 Cl\,1202.7-1224     &  Cl\,1202     & 0.4244  &  518 $_{-104}^{+92}$ &19\\ 
 Cl\,1059.2-1253     &  Cl\,1059     & 0.4561  &  510 $_{-56}^{+52}$  &41\\ 
 Cl\,1054.7-1245     &  Cl\,1054-12  & 0.7498  &  504 $_{-65}^{+113}$ &36\\ 
 Cl\,1018.8-1211     &  Cl\,1018     & 0.4732  &  486 $_{-63}^{+59}$  &33\\ 
 Cl\,1040.7-1155     &  Cl\,1040     & 0.7043  &  418 $_{-46}^{+55}$  &30\\ 
 Cl\,1420.3-1236     &  Cl\,1420     & 0.4959  &  218 $_{-50}^{+43}$  &24\\ 
 Cl\,1119.3-1129     &  Cl\,1119     & 0.5500  &  166 $_{-29}^{+27}$  &17\\
\tableline
\end{tabular}
}
\tablecomments{Col. (1): Cluster name.  Col. (2): Short cluster name. 
Col. (3) Cluster redshift. Col. (4) Cluster velocity dispersion. 
Col. (5) Number of spectroscopic members.
Redshifts and velocity dispersions are taken from Halliday et al. (2004) and
Milvang-Jensen et al. (2008). 
}
\end{center}
\end{table}

\section{The definition of the various environments}

As can be seen in Table~1, EDisCS structures cover a wide range of
velocity dispersions, from massive clusters to groups. For brevity, we
will refer collectively to these structures as ``EDisCS clusters''.
In addition, within the EDisCS data set it is possible to investigate
the spectroscopic properties of galaxies in even less densely
populated environments, at the same redshift as our main structures.

In a redshift slice within $\pm 0.1$ in $z$ from the cluster/group
targeted in each field, where we are sure the spectroscopic catalog
can be treated as a purely \textit{I}-band selected sample with no
selection bias, we have identified other structures as associations in
redshift space as described in \pog.  These associations have between
3 and 6 galaxies and will hereafter be referred to as ``poor groups''.
We did not attempt to derive velocity dispersions for these systems,
given the small number of redshifts per group. In total, our poor
group sample comprises 84 galaxies brighter than the magnitude limits
adopted for our analysis (absolute V magnitude brighter than -20, see
below).

Finally, within the same redshift slices, any galaxy in the
spectroscopic catalogs that is not a member of our clusters, groups or
poor group associations is treated as a ``field'' galaxy.  This field
galaxy sample is composed of 162 galaxies brighter than our limit and
should be dominated by galaxies in regions less populated and less
dense than the clusters and groups, although will also contain
galaxies belonging to poor structures that went undetected in our
spectroscopic catalog. Our field sample is therefore far from being
similar to the galaxy sample in general ``field'' studies, which will
be dominated by a combination of group, poor group, and field galaxies
according to our environment definition.

The median redshift is 0.58 for the field sample and 0.66 for the poor
group sample.  Redshift and EW([OII]) distributions of our poor group
and field samples are given in \pog.  Computing galaxy masses as
outlined in \S7, we find that the mass distribution of galaxies varies
significantly with environment,
progressively shifting towards higher masses from the field to the
poor groups to the clusters. This corresponds to a difference in the
galaxy luminosity distributions, which was shown in \pog.  We build
field and poor group samples that are matched in mass to the cluster
sample (hereafter the ``mass-matched'' field and poor group samples),
drawing for each cluster galaxy a field or poor group galaxy with a
similar mass.  In the following, we will present the results for both
mass-matched and unmatched samples, but show only mass-matched values
in all figures.

\section{Projected local galaxy densities}

The projected local galaxy density is computed for each
spectroscopically confirmed member of an EDisCS cluster. It is derived
from the circular area $A$ that in projection on the sky encloses the
N closest galaxies brighter than an absolute V magnitude
$M_{lim}^V$. The projected density is then $\Sigma$ = N/$A$ in number
of galaxies per square megaparsec. In the following we use N=10, as
have most previous studies at these redshifts. For about 7\% of the
galaxies in our sample, the circular region containing the 10 nearest
neighbors extends off the chip.  Since the local densities of these
sources suffer from edge effects, they were excluded from our
analysis.

Densities are computed both by adopting a fixed magnitude limit
$M_{lim}^V= -20$ and by letting $M_{lim}^V$ vary with redshift between
-20.5 at $z=0.8$ and -20.1 at $z=0.4$ to account for passive
evolution. Absolute galaxy magnitudes are derived as described in
\pog.  We have used two different radial limits to derive the mean
properties of galaxies in each density bin.  First, we tried using
only galaxies within $R_{200}$ (the radius delimiting a sphere with an
interior mean density of 200 times the critical density, approximately
equal to the cluster virial radius).  We also tried including all
galaxies in our spectroscopic sample, regardless of distance from the
cluster center.  The values of $R_{200}$ computed for our clusters, as
well as sky maps showing $R_{200}$ relative to the extent of the
spectroscopic sample, are given in Poggianti et al. (2006). For most
clusters, our spectroscopy extends to $R_{200}$, while severe
incomplete radial sampling occurs for one cluster, Cl1232, which will
be treated separately when relevant, e.g. in \S7.2. The results do not
change whether we confine our analysis to $R_{200}$ or use our full
spectroscopic sample, nor whether we use a fixed or a 
varying magnitude limit.  
Therefore, to maximise the number of galaxies
we can use and to minimize the statistical errors, we show the results
for $M_{lim}^V= -20$ and with no radial limit, unless otherwise
stated.

We apply three different methods to identify the 10 cluster members
that are closest to each galaxy.  These yield three different
estimates of the projected local density, which we compare in order to
assess the robustness of our results.

In the first method, the density is calculated using all galaxies in
our photometric catalogs and is then corrected using a statistical
background subtraction.  
The number of field galaxies within the circular area $A$ and down to
the magnitude limit adopted for the cluster is estimated from the
$I$-band number counts derived for a 4 deg x 4 deg area by Postman et
al. (1998).

\begin{figure*}[t]
\vspace{-3cm}
\centerline{\hspace{5cm}\includegraphics[width=1.0\columnwidth]{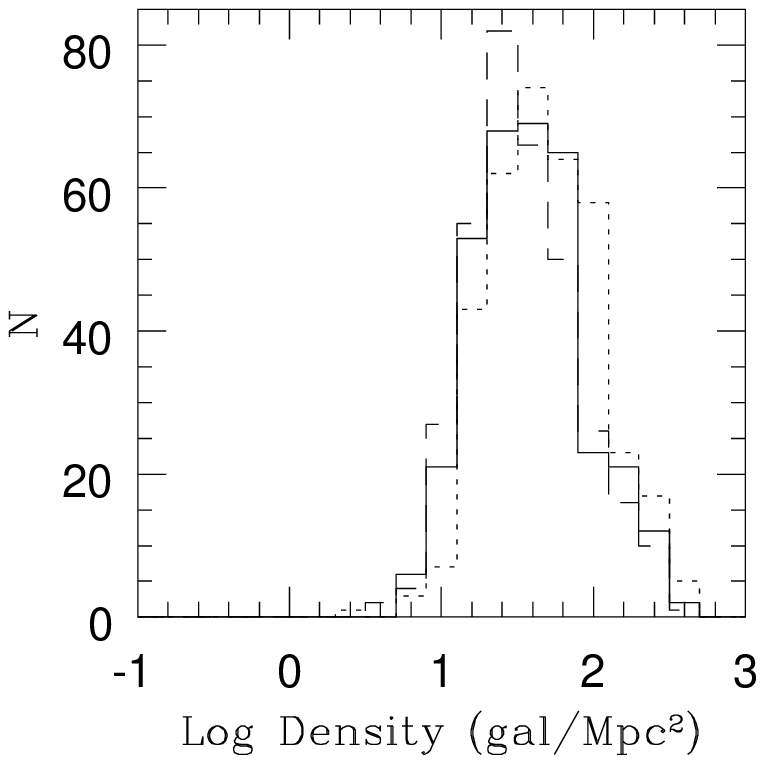}\hfill\hspace{-9cm}\includegraphics[width=1.0\columnwidth]{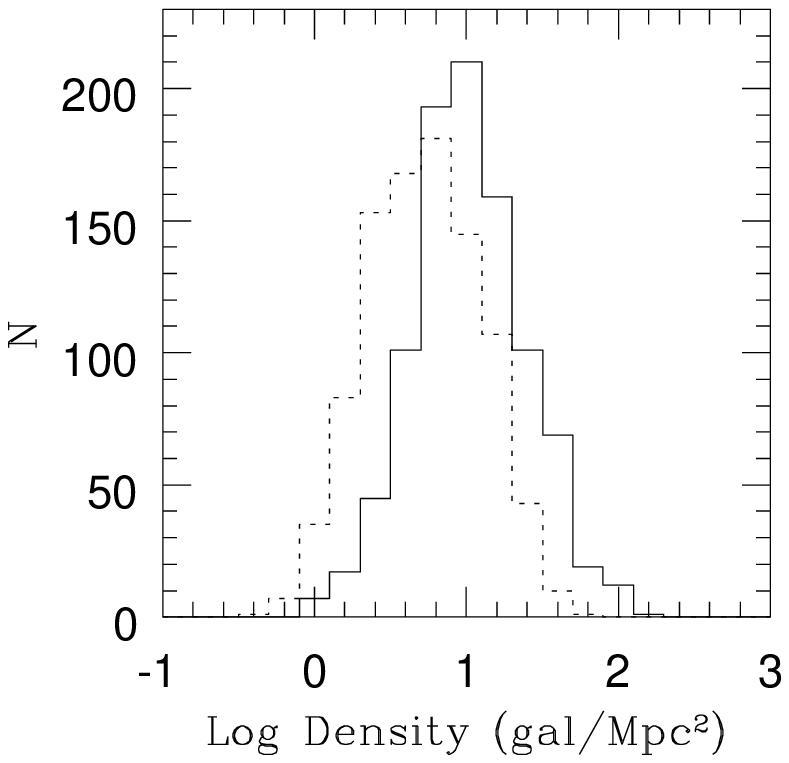}}
 \caption{{\bf Left} Projected local density distribution of
spectroscopic members of EDisCS clusters (\S4).  Three different
estimates are shown (see text for details): using a statistical
background subtraction (short dashed histogram) and using photometric
redshifts retaining photo-z members based on integrated probabilities
(solid histogram) or on estimated photo-z (long dashed histogram).
{\bf Right} Projected local density distribution of spectroscopic
members of low redshift clusters (SDSS) (\S5).  Two different
estimates are shown (see text for details): using a statistical
background subtraction (solid histogram) and using only spectroscopic
members to compute the local density (dashed histogram).
 \label{vdisden}}
 \end{figure*}

In the other two methods we include only those galaxies that are
considered cluster members according to photometric redshift
estimates.  As described in detail in Pell{\'o} et al. (in prep.),
photometric redshifts were computed for EDisCS galaxies using two
independent codes:  a modified version of the publicly available Hyperz
code (Bolzonella, Miralles \& Pell{\'o} 2000), and the code of Rudnick et
al. (2001) with the modifications presented in Rudnick et al. (2003).
We use two different criteria to retain cluster members and reject
probable non-members. In the first case, a galaxy is accepted as
a cluster member if the integrated probability that the galaxy lies
within $\pm 0.1$ in $z$ from the cluster redshift is greater than a
specific threshold for {\sl both} photometric redshift codes. The
probability threshold is chosen to retain about 90\% of the
spectroscopically confirmed members in each cluster. In the other
method a galaxy is retained as a cluster member if the best
photometric estimate of its redshift from the Hyperz code is within
$\pm 0.1$ in $z$ from the cluster redshift.  The projected local
density distributions obtained with the three methods are shown in
Fig.~\ref{vdisden}.

We note that throughout the paper we use only proper (not comoving)
lengths, areas, and volumes.  For example,
our local densities are given as the number of galaxies per $\rm
Mpc^2$, as measured by the rest-frame observer. This choice is
dictated by the fact that with local densities we are investigating
vicinity effects, and gravitation depends on proper distances.  

\section{Low redshift sample}

Using the SDSS, we have compiled a sample of clusters and groups at
$0.04<z<0.1$.  This sample serves as a low-redshift baseline with
which we can compare our high-z results.  The SDSS cluster sample is
described in \pog\footnote{Note that for this work we have excluded
five of the 28 clusters used in \pog (A1559, A116, A1218, A1171 and
A1279) that have less than 10 spectroscopic members within the area
and magnitude limits adopted here.} and comprises 23 Abell clusters
with velocity dispersions between 1150 and 200 $\rm km \, s^{-1}$,
with an average of 35 spectroscopically confirmed members per
cluster. To approximate the EDisCS spectroscopic target selection,
which was carried out at rest-frame $5000 \pm 500$ \AA $\,$,
we used a $g$-selected sample extracted from the SDSS spectroscopic
catalogs.

Local densities were computed for spectroscopic cluster members
(within 3$\sigma$ from the cluster redshift) that lie within $R_{200}$
from the cluster center. For Sloan, the radial cut is necessary to
approximate the EDisCS areal coverage, which reaches out to about
$R_{200}$. We choose a galaxy magnitude limit of $M_V<-19.8$, which
maximizes the number of galaxies we can use in our analysis and would
correspond to $-20.1$ at $z=0.4$ and $-20.5$ at $z=0.8$ under the
assumption of passive evolution.

As for EDisCS, local densities were derived using the circular area
encompassing the 10 nearest neighbors. Two methods were employed to
obtain two independent estimates of local densities.  In the first
method, we find the distance to the tenth-nearest projected neighbor
considering only spectroscopically confirmed members brighter than
$M_V<-19.8$. In the second method, we include the 10 nearest projected
neighbors that are within the Sloan photometric catalog and that have
an estimated absolute magnitude satisfying $M_V<-19.8$.  Absolute
magnitudes were derived from observed magnitudes assuming that all
galaxies lie at the cluster redshift and using the transformations of
Blanton et al. (2003). Both the spectroscopic and the photometric
densities were computed from catalogs of galaxies located over an area
much larger than $R_{200}$ to avoid edge effects.

Since the spectroscopic completeness within $R_{200}$ of the SDSS
clusters is on average about 84\%, the first method is likely to
underestimate the ``true'' density.
Using only the photometric catalog, we ignore the fact that some of
the 10 galaxies might be in the background or foreground of the
cluster.  The second method therefore overestimates the value of the
density. Thus, the spectroscopically-based and photometrically-based
local densities represent lower and upper limits to the local density.
Density values that have been corrected for spectroscopic
incompleteness and for background contamination
will lie between these two values.  

The local density distributions derived with the two methods are
shown in the right panel of Fig.~\ref{vdisden}. 
As expected, the distribution
of spectroscopically-based densities is shifted to slightly lower
densities than the distribution based on photometry, 
but in the following we will see that the conclusions reached by the two
methods are fully consistent.
Compared to the density distribution at high-z shown in the left panel
of the same figure, the low-z distribution is shifted to lower densities.
In Poggianti et al. (in prep.) we show this is due to the fact that
high-z clusters are on average denser by a factor of
$(1+z)^3$ compared to nearby clusters, with possible strong
consequences on galaxy evolution. 

The values of EW(\oii) measured with the method we used for EDisCS
spectra are in very good agreement with those measured by Brinchmann
et al.  (2004a,b) for star-forming galaxies in the SDSS (see Fig.~3 in
Poggianti et al.~(2006)).  However, for EDisCS galaxies, the EW(\oii)
was measured only when the line was present in emission, and a value
EW(\oii)=0 was assigned when no line was present. In addition, each 1D
and 2D spectrum was visually inspected.  In contrast, the SDSS \oii
measurements of Brinchmann et al. (2004a,b) are fully automated and
can even yield a (small) value in absorption that is compatible with 0
within the error and that cannot be ascribed to the {\oii}3727 line.
To take this into account, for SDSS clusters we have used the EWs
provided by Brinchmann et al. (2004a,b), but have forced the EW(\oii)
to be equal to 0 when the value provided by Brinchmann et al. is
$EW<0.8$ \AA $\,$ in emission. Moreover, to be compatible with EDisCS,
we do not exclude AGNs from our SDSS analysis. Finally, the redshift
range of our Sloan clusters was chosen as a compromise to minimize
aperture effects while still sampling sufficiently deep into the
galaxy luminosity function. The $3\arcsec$ SDSS fiber diameter covers
the central 2.4-5.4 kpc of galaxies depending on redshift, compared to
the $1\arcsec$ EDisCS slit covering 5.4-7.5 kpc at high redshift. In
the following, we assume that [OII] equivalent widths do not change
significantly over these different areas.

\section{Results}

\subsection{\oii strength and star-forming fractions as a function of local density at high-z}

We first investigate how the strength of the \oii line varies in
EDisCS clusters as a function of projected local density.
Figure~\ref{main} shows the mean equivalent width of \oii measured
over all galaxies that are spectroscopically confirmed members of
EDisCS clusters (black symbols), in bins of local density.  The three
different estimates of density described in \S3 (empty and filled
circles and crosses) yield similar results within the errors. In the
following, errors on mean equivalent widths are computed as bootstrap
standard deviations, and errors on fractions are computed from
Poissonian statistics.

The mean EW computed over all galaxies is consistent with being flat
up to a density $\sim 70 \,\rm gal/Mpc^2$ (the Kendall's probability
for an anticorrelation in the three lowest density bins
is only 40\%), then decreases at higher
densities. These trends arise from a combination of the incidence of
star-forming galaxies and the relation between density and EW in
star-forming galaxies.

As shown in fig.~\ref{fractions}, the proportion of star-forming
galaxies tends to decline at higher density.  
The fraction of star-forming galaxies decreases
from about 60\% to $\le$30\%.
Using the average of
the values given by the 3 membership methods, the Kendall test gives a
95\% probability of an anticorrelation. 

In contrast, fig.~\ref{oiionly} shows that the mean EW(\oii) computed
{\it only for galaxies with emission lines} does not correlate with
local density (the Kendall's correlation probability is 38\%). It is
consistent with being flat over most of the density range, except for
the highest density bin centered on $\sim 450$ galaxies per $\rm
Mpc^2$, where it drops by a factor 2 to 3. As will be shown in \S8,
the highest density bin is populated only by elliptical galaxies,
whose weak [OII] may be related to the presence of an AGN. It is thus
not surprising that this bin stands out from the other bins, where
star-forming spirals dominate the mean behavior.

The constancy of the average \oii equivalent width 
in star-forming galaxies at most densities 
suggests that 
as long as star formation is active, 
it is on average unaffected by local environment, at least in
clusters.

\subsection{Comparison with poor groups and the field at $z=0.4-0.8$}

The mean EW(\oii) among all of our field galaxies is 10.7$\pm$1.5 \AA
$\,$ for the mass-matched sample (dashed lines in fig.~\ref{main}),
and 13.5$\pm$1.5 \AA $\,$ for
the unmatched sample.  Similar
values are found for poor group galaxies: 
11.6$\pm$1.8 \AA $\,$ for the mass-matched sample (dotted lines)
and 14.5$\pm$1.8 \AA $\,$ for the unmatched sample.
These values are comparable within the uncertainties to the 
values measured in the low density regions of the clusters.

Similarly, when considering only galaxies with ongoing
star formation, the mass-matched field and poor group values
are 
comparable to those at most densities in clusters
 (fig.~\ref{oiionly}). The mean EW of \oii field and
poor group galaxies is 17.2$\pm$1.5 \AA $\,$ and 14.5$\pm$2.2 \AA $\,$, 
respectively
(18.0$\pm$1.5\AA $\,$ and 18.9$\pm$2.2\AA $\,$ in the unmatched samples). 

Finally, the fraction of \oii galaxies in the field is 62$\pm8$\%
(matched, fig.~\ref{fractions}) 
and 72$\pm8$\% (unmatched).  In the poor groups, the
mass-matched \oii fraction is 80.0 $\pm10$\% (matched) and 77$\pm10$\%
(unmatched).  The star-forming fraction in
the field is compatible with that
observed at most densities in clusters,
while the poor group fraction is slightly higher.

Therefore, relative to clusters, the unmatched
poor groups and field have higher average
EWs and star-forming fractions.  Our results indicate that this is
primarily due to differences in the galaxy mass distribution with
environment.  Using galaxy samples with similar mass distributions, we
find that the EW properties of star-forming galaxies do not differ 
significantly between
clusters, poor groups, and the field, 
neither with local density within clusters as shown in the previous section.

\subsection{\oii-local density relation as a function of cluster mass}

To assess whether the relation between star formation and density is
the same in structures of different mass, we divide our cluster sample
into different velocity dispersion bins and show the 
correlation found above, between the star-forming fraction and density,
in Fig.\ref{fraeach}. The analysis is now done using only galaxies
within $R_{200}$. As above, errors on fractions are
computed from Poissonian statistics.

The two most massive clusters ($\sigma > 800 \, \rm km \, s^{-1}$ )
exhibit a flatter relation. i.e. have lower [OII] fractions in the
three lowest density bins, than clusters with $\sigma < 800 \, \rm km
\, s^{-1}$.\footnote{$\sigma > 800 \, \rm km \, s^{-1}$ corresponds to
the threshold between less(more) massive clusters with an average
total cluster [OII] fraction above(below) 50\% (\pog).}  

In contrast, we find that at low-z this relation is indistinguishable
in clusters with $\sigma$ above and below $800 \, \rm km \, s^{-1}$,
in agreement with previous works that found no dependence of the
correlation between the star-forming ($\rm H\alpha$-emitting) fraction
and density from the cluster velocity dispersion in the local universe
(Lewis et al. 2002, Balogh et al. 2004).

We note that although the relation between [OII] fraction and local
density varies with cluster mass at high-z, the relation between star formation
and {\it physical} three-dimensional space density may be constant,
since the distribution of physical densities in each projected 2D
density bin varies with cluster mass (Poggianti et al. in prep.).

Dividing the high-z sample into finer velocity dispersion bins, we do
not find a continuous trend with velocity dispersion (Fig.\ref{fraeach}).  
Systems with
$600<\sigma<800 \, \rm km \, s^{-1}$ may lie at larger or comparable
[OII] fractions than systems with $\sigma<600 \, \rm km \, s^{-1}$,
but our errorbars are too large to draw any conclusion.

In \pog we studied how the fraction of [OII] galaxies depends on the
cluster velocity dispersion $\sigma$.  In an [OII] fraction-$\sigma$
diagram, one can identify three groups of structures: a) high-mass
structures, all with low [OII] fractions; b) low-mass structures with
{\it high} [OII] fractions and c) low-mass structures with {\it low}
[OII] fractions (the so-called ``outliers'' in \pog).  In addition to
having a low [OII] fraction, the outliers have a low fraction of blue
galaxies (De Lucia et al.~2007), a high fraction of early-type
galaxies given the measured velocity dispersions (Simard et al. 2008),
and peculiar [OII] equivalent width distributions (\pog).  Hence,
galaxies in the outliers resemble those in the cores of much more
massive clusters.

The presence of low-mass structures with low [OII] fractions
in our sample could be responsible for a non-monotonic trend of the
[OII] fraction-density relation with cluster mass.  We study the
dependence of the [OII] fraction on local projected density for the
three groups separately in the right panel of Fig.\ref{fraeach}.

Except for the lowest density bin where the results of all three
groups are compatible within the errors, the trend with local density
is different in the three groups. At any given density, the
star-forming fraction in low-mass, low-[OII] groups is significantly
lower than those in low-mass high-[OII] systems.

In \pog we found that the global [OII] fraction in distant clusters
relates to the system mass, but not {\it solely} to the system mass,
at least as estimated from the observed velocity dispersion. Here we
find that the [OII] fraction does not depend solely on projected local
density but also on global environment, and that variations in the
star-forming fraction-density relation do not depend uniquely on
cluster mass. In principle, the correlations with system mass and
local density could have a single common origin, 
from i.e. a correlation between galaxy properties and physical
density in three-dimensional space.  In a separate paper (Poggianti et
al.~in prep.), we use numerical simulations to investigate the
relations between projected local density, physical 3D density, and
cluster mass.  The aim of that paper is a simultaneous interpretation
of the observed trends with local density and cluster mass presented
in this paper and in \pog.

\subsection{The EW([OII])-density relation at low redshift}

\begin{figure}
\vspace{-8cm}
\centerline{\hspace{2cm}\includegraphics[width=16cm]{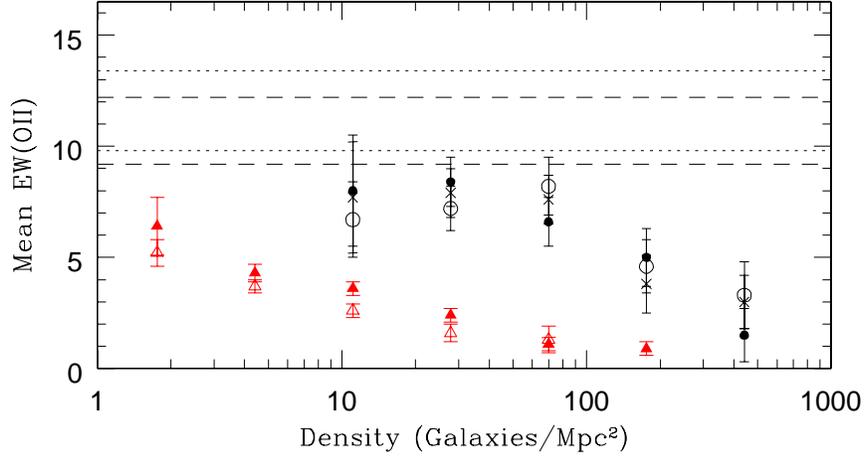}}
\caption{Mean equivalent width of \oii versus projected local density.
The mean EW(\oii) is computed over all galaxies. The high redshift
points (EDisCS), described in \S6.1,
 are shown in black as empty circles (statistical
subtraction), filled circles (photo-z probable members) and crosses
(photo-z within $\pm 0.1$ from $z_{clu}$). The low redshift points
(SDSS), described in \S6.4,
are shown in red as empty triangles (density computed using
only spectroscopic members) and filled triangles (full photometric catalog).  
Errors are computed as bootstrap standard deviations
from the mean using 100 realizations.  The horizontal dashed and
dotted lines delimit the values found in the field and in poor groups
at high redshift using mass-matched samples (see text).
 \label{main}}
 \end{figure}

\begin{figure}
\vspace{-8cm}
\centerline{\hspace{2cm}\includegraphics[width=16cm]{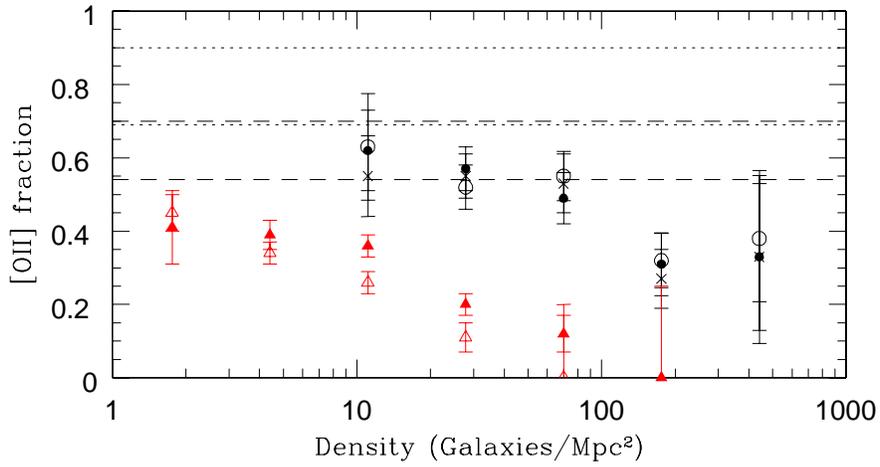}}
 \caption{Fraction of galaxies with \oii in emission versus local density.
Symbols as in fig.~\ref{main}. Black symbols (circles) are 
EDisCS points at $z=0.4-0.8$ (\S6.1).
Red symbols (triangles) are SDSS points at low redshift (\S6.4). Errors on data points
are computed from Poissonian statistics. The horizontal dashed and dotted 
lines delimit the value for field and poor group galaxies at $z=0.4-0.8$. 
 \label{fractions}}
 \end{figure}

\begin{figure}
\vspace{-8cm}
\centerline{\hspace{2cm}\includegraphics[width=16cm]{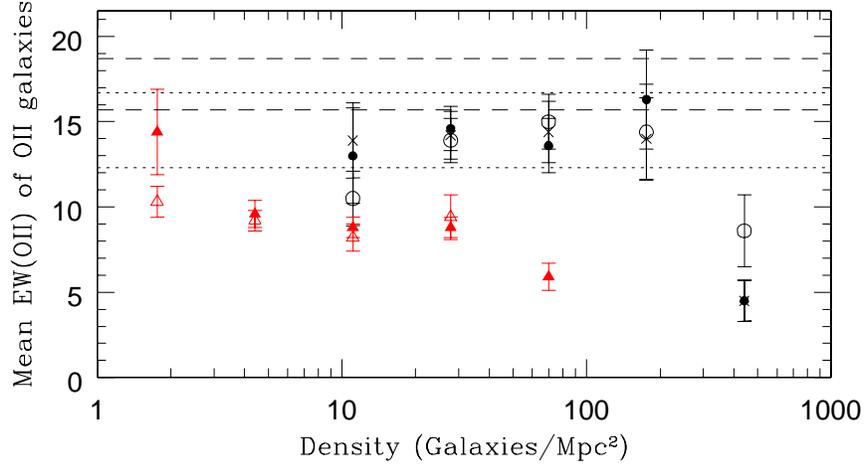}}
 \caption{The mean EW(\oii) only among
galaxies with \oii in emission (EW$>3$ \AA). Symbols as in fig.~\ref{main}.
Black symbols (circles) are EDisCS points at $z=0.4-0.8$ (\S6.1).
Red symbols (triangles) are SDSS points at low redshift (\S6.4). 
Errors are computed as bootstrap standard deviations. 
The horizontal dashed and dotted 
lines delimit the value for mass-matched
field and poor group galaxy samples at $z=0.4-0.8$. 
 \label{oiionly}}
 \end{figure}

\begin{figure*}
\vspace{-8cm}
\centerline{\hspace{2cm}\includegraphics[width=0.7\columnwidth]{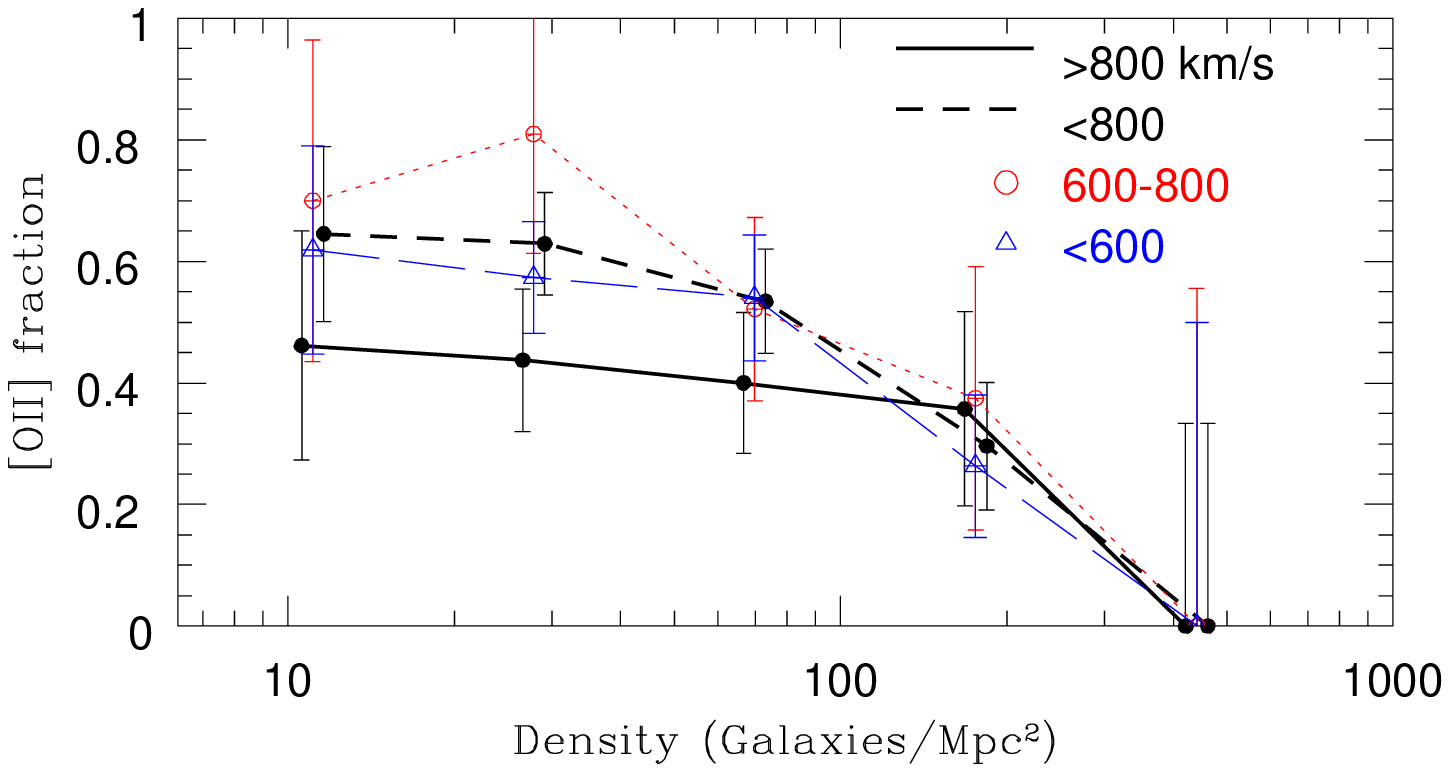}\hfill\hspace{-3.5cm}\includegraphics[width=0.7\columnwidth]{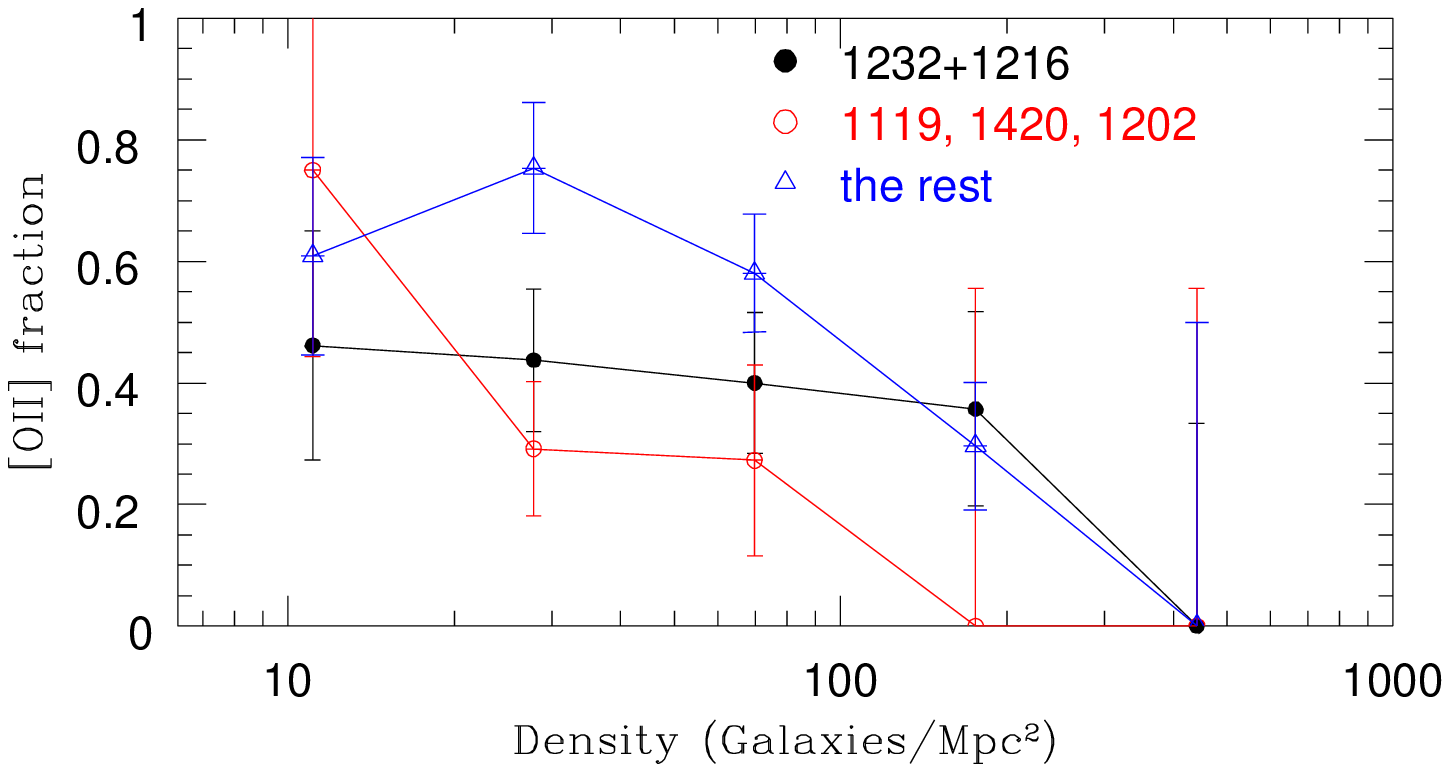}}
 \caption{Fraction of galaxies with \oii in emission versus local
density for different subsets of our cluster sample: {\bf Left}
Clusters in different velocity dispersion bins: $\sigma > 800 \, \rm
km \, s^{-1}$ (thick solid line) and $\sigma < 800 \, \rm km \,
s^{-1}$ (thick dashed line). A small shift in density has been applied
to allow a better visibility of the errors. The red dotted lines
indicate clusters with $600 < \sigma < 800$ km s$^{-1}$ and
blue, long-dashed lines respresent systems with $\sigma < 600$ km s$^{-1}$.  {\bf Right} The two most massive clusters ($\sigma > 800$
km s$^{-1}$, cl1232 and cl1216); the three outliers in the
[OII]-$\sigma$ relation (cl1119, cl1420 and cl1202), and all the
remaining clusters.  Densities have been computed using the
high-probability photometric-redshift membership (sec.~4).
 \label{fraeach}}
 \end{figure*}

The SDSS results are shown as red symbols (triangles) in
fig.~\ref{main}, fig.~\ref{fractions} and fig.~\ref{oiionly}.  At
low-z, the mean EW(\oii) computed for all galaxies continuously and
smoothly decreases with local density (fig.~\ref{main}).  This trend
is driven by the decrease in the star-forming fraction with local
density (fig.~\ref{fractions}).
Using the average of the values obtained with
the two density estimates, the Kendall's test yields a 98.5\%
probability of an anti-correlation.

As at high-z, the mean \oii strength of \oii-galaxies does not vary
significantly over most of the density range (Kendall's probability
82.6\%), except for a decrease in the highest density bin
(fig.~\ref{oiionly}). These findings are in agreement with previous
low-z results based on SDSS and 2dFGRS (Lewis et al. 2002, Balogh et
al. 2004).

From a quantitative point of view, blindly comparing the high-z and
the low-z results, at any projected density in common we observe a
lower average EW([OII]) at low-z.  This is due to both a lower average [OII]
fraction and a lower average EW([OII]) in star-forming galaxies at low-z. 
Taken at face value, this indicates that both the proportion
of star-forming galaxies and the star formation activity in them
decrease with time at a given density. However, it is
worth stressing that observing similar projected densities at
different redshifts does not imply similar physical densities,
since the correlation between projected and 3D density varies with
redshift (Poggianti et al. in prep.). Hence, a quantitative
comparison of results at different epochs at a given projected density
cannot be interpreted as a direct measure of the decline of the star
formation activity with time for similar ``environmental'' physical
conditions.

Moreover, we stress again that to the low-z density distribution is
shifted to lower densities compared to clusters at high-z.  This is
due to the fact that high-z clusters are on average denser by a factor
of $(1+z)^3$ compared to nearby clusters, as discussed in Poggianti et
al. (in prep.).

From a qualitative point of view, the only difference between high-
and low-z observations is the fact that the high-z EW(OII) averaged
over all galaxies is consistent with being flat up to a density $\sim 70 \,\rm
gal/Mpc^2$ (40\% probability for an anticorrelation), 
while the low-z trend smoothly declines towards higher
densities with no discontinuity (98.5\% probability for an anticorrelation).  

For the rest, the trends of \oii
equivalent widths and star-forming fraction with local density are
qualitatively very similar at $z=0$ and $z=0.8$, showing that 
an [OII] fraction-density relation similar to that observed locally is 
already established in clusters at these redshifts, and that the
activity in star-forming cluster galaxies, when assessed from the
EW of the [OII] line, does not appear to 
depend strongly on local density at any redshift.

\section{Star formation rates}

The star formation rate (SFR) of a galaxy can be roughly estimated
from the [OII] line flux in its integrated spectrum. The equivalent
width is the ratio between the line flux and the value of the
underlying continuum.  Thus, it is not directly proportional to the
star formation rate. For example, a faint late-type galaxy in the
local Universe usually has a higher equivalent width, but a comparable
or lower star formation rate than a more luminous spiral.

For this reason, the analysis presented above does not yield
information on absolute star formation rates, but only on the
strength of star formation relative to the galaxy $\sim U$ rest-frame
luminosity, which itself depends on the current star formation
activity.

We have derived star formation rates of galaxies with EDisCS spectra
by multiplying the value of the observed equivalent width by the value
of the continuum flux estimated from our broad-band photometry. For
the latter we have used the total galaxy magnitude as estimated from
spectral energy distribution (SED) fitting (Rudnick et al. 2008),
assuming that stellar population differences between the galactic
regions falling in and out of the slit are negligible (see
\S5).\footnote{We do not attempt to compare with SFRs in Sloan, as
SFRs are more sensitive to aperture effects than EWs, and
dishomogeneity in observations and photometry between the two datasets
would render a quantitative comparison highly uncertain.}

We use the conversion $SFR(M_{\odot}yr^{-1}) = L([OII]) \, erg \,
s^{-1}/(1.26 \times 10^{41})$ (Kewley et al. 2004), adopting an
intrinsic (with no dust attenuation) flux ratio of [OII] and $\rm
H\alpha$ equal to unity with no strong dependence on metallicity, as
found by Moustakas et al. (2006).  At this stage we do not attempt to
correct our star formation estimates for dust extinction. Locally, the
typical extinction of [OII] relative to $\rm H\alpha$ is a factor 2.5,
and the typical extinction at $\rm H\alpha$ is an additional factor
2.5-3, so our SFR estimates would be corrected by a factor $\sim 7$
for extinction,\footnote{A comparison of our [OII]-based SFRs with
those derived from $\rm H\alpha$ narrow-band photometry from Finn et
al. (2005) shows that the relation between the two does not strongly
deviate from that derived using the local typical factor $\sim 7$ for
extinction.}  but there are large galaxy-to-galaxy and redshift
variations and they are hard to derive using only optical spectra.
Dust-free SFRs based on Spitzer data of the EDisCS clusters will be
presented in Finn et al. (in prep).

Adopting the same criteria and galaxy sample as in Sec. 4, we derive
the mean star formation rate for galaxies in bins of local projected
density and compute errors as bootstrap standard deviations.

\begin{figure}[t]
\vspace{-4cm}
\centerline{\hspace{2cm}\includegraphics[width=16cm]{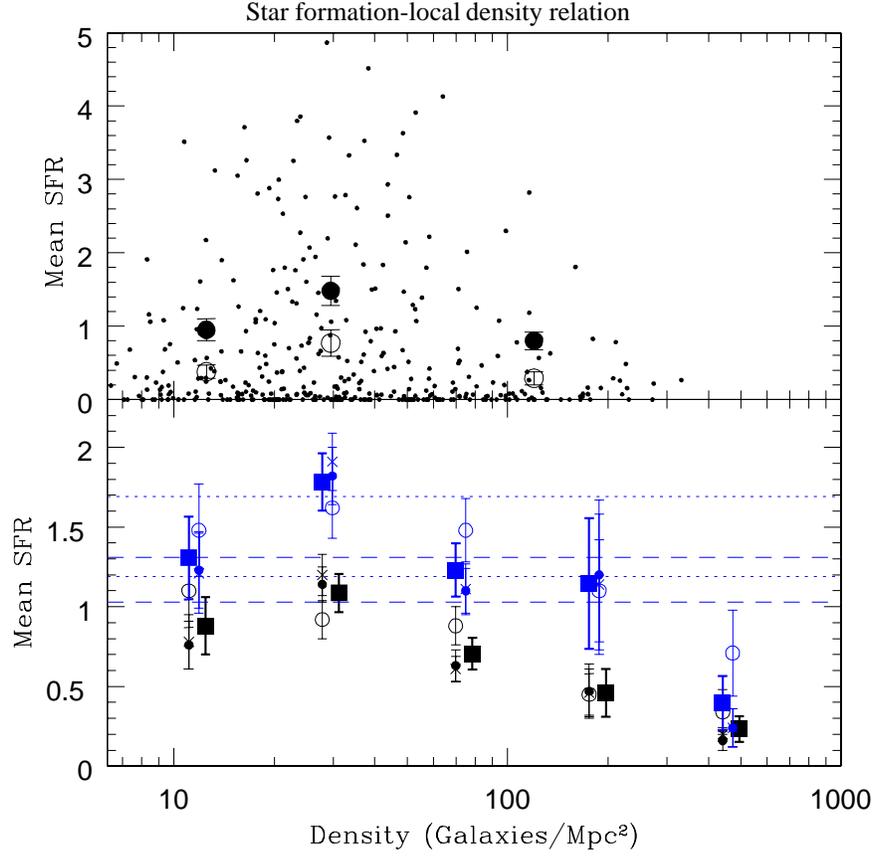}}
 \caption{{\bf Bottom} Mean SFR in solar masses per year for galaxies
in different density bins. All galaxies: black symbols. Only
star-forming galaxies: blue symbols. Different symbols indicate
the values obtained for the 3 membership criteria, as in
Fig.~\ref{main}.  Large squares represent the average of the values for
the 3 membership methods. A small shift around the center of each
density bin has been applied to the different points to avoid
confusion.  The dashed and dotted lines delimit the 1$\sigma$ error
around the value for field and group galaxies, respectively. For
clarity, only the star-forming field and poor group values for the
mass-matched samples are shown.  {\bf Top} SFR in solar masses per
year for all individual galaxies. The mean (filled circles) and median
(empty circles) SFR in star-forming galaxies
are shown for three equally populated density bins
(see text).
\label{mainsfr}}
\end{figure}

The lower panel of Fig.~\ref{mainsfr} shows the results including all
cluster members as black symbols. The mean SFR is $\sim 1-1.2$
$M_{\odot}yr^{-1}$ in low-density regions and 
declines towards denser regions. The mean SFR might present a maximum
at a density between 15 and 40 \rm galaxies/$\rm Mpc^2$, though within
the errors the values of the two lowest density bins may be
consistent.  The corresponding mean SFR for all galaxies in the field
and poor groups is 0.8-1.2 (0.9-1.3 in the unmatched samples),
comparable to the average in cluster low-density regions.

Considering only star-forming galaxies, the trend with local density
in clusters remains similar and is shifted to higher SFR values with a
maximum of $\sim 1.8$ ${\rm M}_{\odot} {\rm yr}^{-1}$ between 15 and
40 \rm galaxies/$\rm Mpc^2$ (blue points in Fig.\ref{mainsfr}).
Errorbars are larger here due to the reduced number of galaxies.
Nevertheless, the presence of a peak is hinted at by the data at the
1-2$\sigma$ level. To further assess the significance of the peak, we
have computed mean and median SFR values for star-forming galaxies in
5, 4, and 3 equally-populated density bins. The latter are shown in the
top panel of Fig.\ref{mainsfr}, together with the SFR values for 
individual galaxies.

The equally-populated bins confirm the presence of the peak at the 2
to 4$\sigma$ level.  A KS test rejects the null hypothesis of similar
SFR distributions in star-forming galaxies in the peak density bin and
in each one of the other bins with a 98.4\% and 98.2\% probability.
The galaxy mass distribution varies only slightly from one density bin
to another. In any case, we have verified that the significance of the
peak in the mean and median SFR remains the same when matching the
mass distribution of galaxies in the lowest and highest density bin to
that in the bin with the peak.  The distribution of individual points
in the top panel of Fig. \ref{mainsfr} is also visually consistent
with higher SFRs for a significant number of galaxies at densities
between 15 and 50 galaxies/$\rm Mpc^2$.

The peak SFR is higher than the mean values of $1.17\pm0.14$
$M_{\odot}yr^{-1}$ for mass-matched field star-forming galaxies, but
is compatible within the errors with the mass-matched
poor group value of $1.44\pm0.25$
(blue lines in Fig.\ref{mainsfr}). 
The unmatched samples yield similar results ($1.2-1.3$).

\begin{figure}[t]
\vspace{-9cm}
\centerline{\hspace{2cm}\includegraphics[width=16cm]{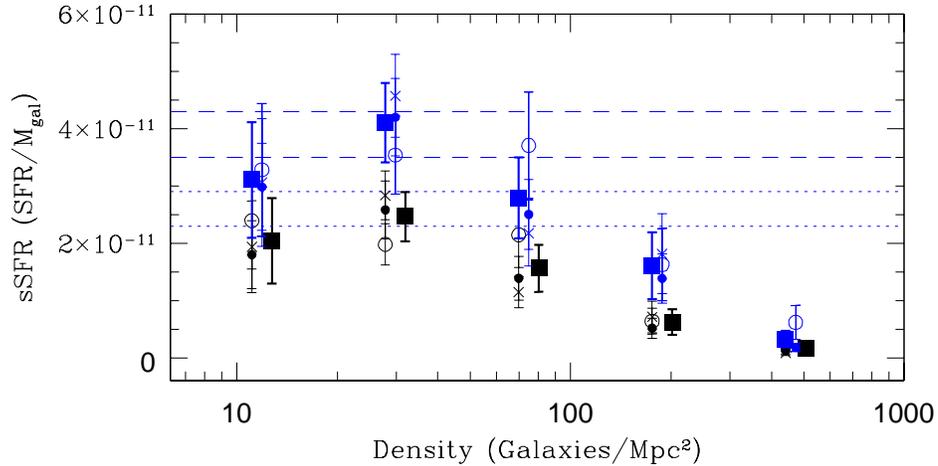}}
 \caption{Mean specific SFR (SFR/M, where M is the galaxy stellar
mass) in $\rm yr^{-1}$ for galaxies in different density bins. All
galaxies: black symbols. Only star-forming galaxies: blue
symbols. Different symbols indicate the values obtained for
the 3 membership criteria, as in Fig.~\ref{main}. Large squares
represent the average of the values for the 3 membership methods. A
small shift around the center of each density bin has been applied to
the different points to avoid confusion.  The dashed and dotted lines
delimit the 1$\sigma$ error around the value for field and group
galaxies, respectively. For clarity, only the star-forming 
field and poor group values for the mass-matched samples are shown.
 \label{mainssfr}}
 \end{figure}

Figure \ref{mainssfr} shows the average {\it specific} star formation
rate (sSFR), defined as the SFR per unit of galaxy stellar mass, as a
function of local galaxy density.
Galaxy stellar masses were computed from rest-frame absolute
photometry derived from SED fitting (Rudnick et al. 2008), adopting
the calibrations of Bell \& De Jong (2001), which are based on a diet
Salpeter IMF.
Cluster trends are similar to the SFR-density diagram, reinforcing the
picture of a peak and a declining trend on both sides of the peak.
The average specific star formation rates in mass-matched samples of
star-forming field galaxies ($3.9\pm0.4 \, 10^{-11}$ yr$^{-1}$) and of poor
groups ($2.6\pm0.3 \, 10^{-11}$ yr$^{-1}$) are comparable to those found in
the low-density regions of clusters.

Interestingly, using the unmatched samples, the field would be
markedly distinct from the other environments, having higher specific
star formation rates by a factor of two or more.  This shows that on
average our star-forming field galaxies are forming stars at more than
twice the rate per unit of galaxy mass than star-forming galaxies in
any other environment we have observed, and that this is due to their
average lower galaxy mass.

Our results show that in distant clusters the average star formation
rate and the specific star formation rate per galaxy, computed both
over all galaxies and only among star-forming galaxies, may not follow
a continuously declining trend with density. The most striking result
is the significance of the peak in the SFR of star-forming galaxies
discussed above.
The average star formation rate over all galaxies decreases with
density in the general field at $z=0$ (Gomez et al. 2003), but distant field
studies have found that the relation between average star formation
rate over all galaxies and local density was reversed at $z=1$, when
the SFR {\it increases} with density, at least up
to a critical density above which it may decrease again (Cooper et
al. 2007, Elbaz et al. 2007).

These high-z surveys sample different regions of the Universe (the
general ``field'') and slightly higher redshifts than our survey ($z
\sim 0.75 - 1.2$). The range of projected densities in these studies
is likely to overlap with our range only in their highest density
bins, but a direct comparison is hampered by the different measurement
methods of local density. It is compelling, however, that both we and
these studies find a possible peak plus a possible decline on either
side of the peak. Unfortunately, none of these studies sample a
sufficiently broad density range to be sure of the overall trend. It
is possible that the SFR per galaxy at redshifts approaching 1 presents
a maximum at intermediate densities (corresponding to the
groups/filaments that are common to all of these studies), and
declines both towards higher and lower density regions.  Large surveys
sampling homogeneously a wide range of environments and local
densities at $z=0.5-1$ should be able to address this question.

\subsection{Comparing the SFR-density and the EW-density relations}

To summarize the results presented in the previous sections, there are
some notable differences between the ``star formation-density''
relation as depicted by the observed equivalent widths
(EW([OII])-density relation), and that portrayed by the measured star
formation rates (SFR-density relation).  The main differences are best
seen by comparing Fig.~\ref{oiionly} with Fig.~\ref{mainsfr}, and can be
described as follows.

The ``strength'' of star formation in star-forming galaxies, when
assessed from the EW([OII]), is consistent with being
flat with density in clusters (except for the strong
depression in ellipticals in the densest regions), and
to be rather similar in equally massive field and poor group galaxies.  The
``strength'' of star formation in star-forming galaxies, when
represented by the SFR, possibly peaks in clusters at $\sim 30$ galaxies/$\rm
Mpc^2$, exceeding the field value.
This finding appears robust to any statistical test we have applied.
However, data for larger galaxy samples
will be needed to confirm this result.

From the EW([OII])-density relation one would conclude that on average
the star formation activity in currently star-forming galaxies is
invariant with both local and global environment, while from the
SFR-density relation one may conclude that the star formation rate is
possibly boosted by the impact with the cluster outskirts, as several
studies have suggested (see e.g. Milvang-Jensen et al. 2003, Bamford
et al. 2005, Moran et al. 2005). Variations in star formation
histories and dust extinction with density must play a role in causing
the differences between the EW and SFR trends, and may conspire to
keep the EW([OII]) relation flat.  We have instead verified that
variations in the galaxy mass distributions are not responsible for
the SFR peak (see above).

The relation between line EW and local density
is often considered 
equivalent to the SFR-density relation, but we have shown here that
they provide different views of the dependence of the star formation
activity on environment.

\subsection{Cluster integrated SFRs}

We derive cluster-integrated SFRs by summing up the SFRs of individual
galaxies within the projected $R_{200}$.  We derive the individual
SFRs from the [OII] line flux as described in the previous section,
and weight each galaxy for spectroscopic incompleteness as outlined
in \S2.
We do not attempt to
extrapolate to galaxy magnitude limits fainter than the spectroscopic
limit adopted for this paper, thus SFRs in galaxies fainter than
$M_V=-20$ are not included in our estimate.

\begin{figure*}[t]
\vspace{-2cm}
\centerline{\hspace{1cm}\includegraphics[width=0.6\columnwidth]{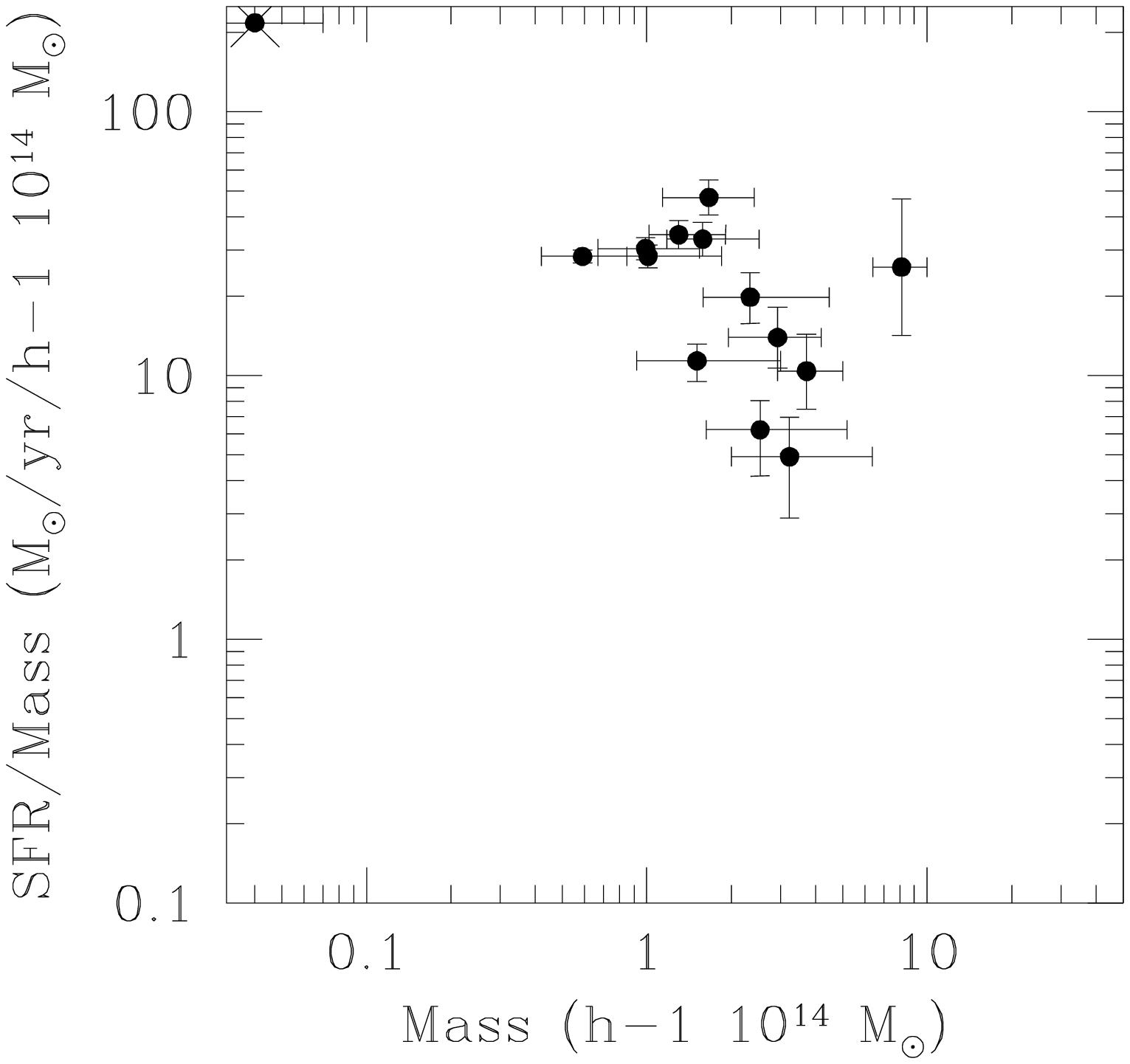}\hfill\hspace{-2cm}\includegraphics[width=0.6\columnwidth]{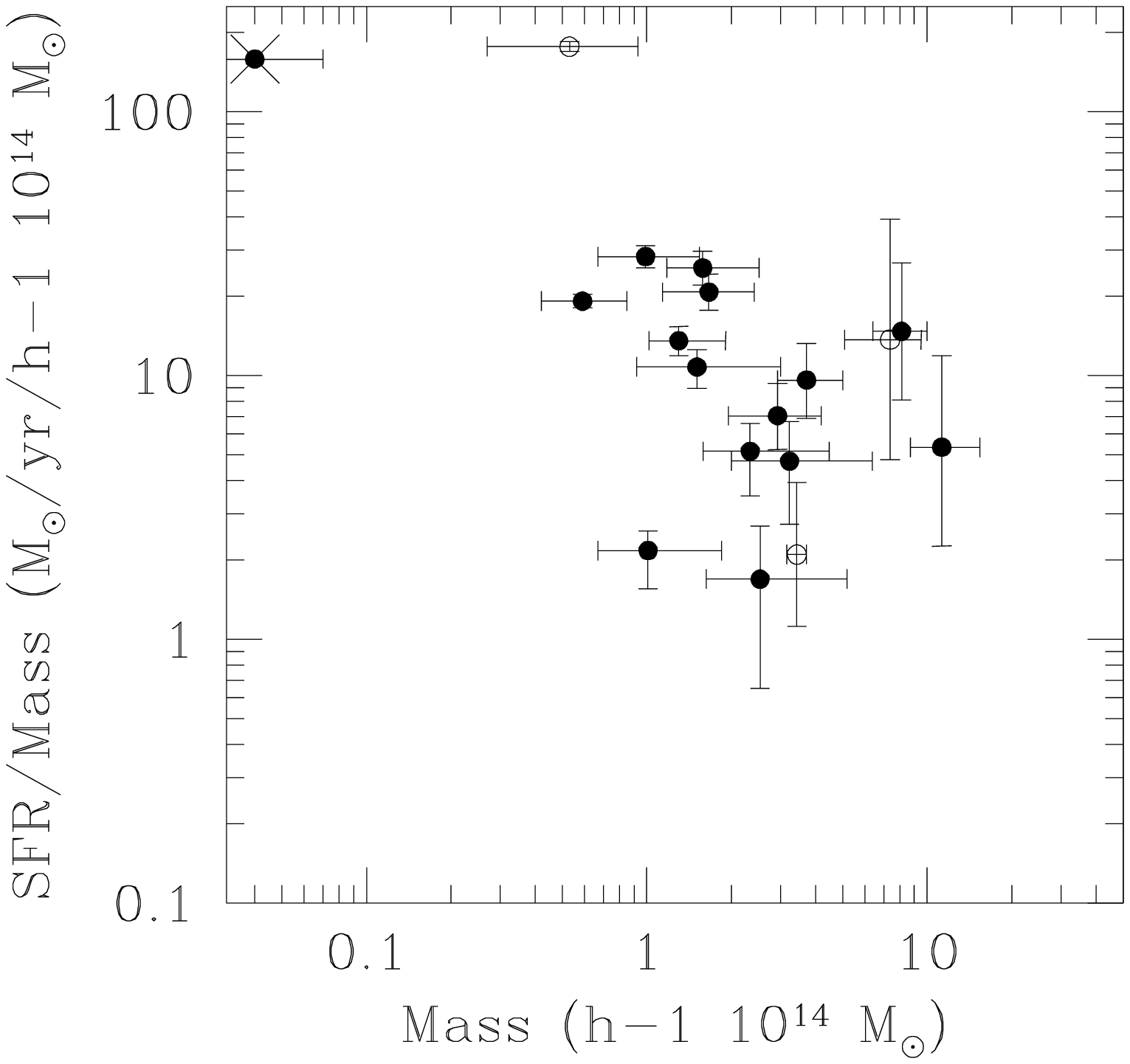}}
 \caption{Integrated cluster SFR per unit of cluster mass plotted as a
function of cluster mass. The large cross identifies our lowest
velocity dispersion group (Cl1119) whose SFR/Mass estimate is unreliable
(see text).  {\bf Left} EDisCS clusters over a radius equal to
$R_{200}$. Cl\,1232 has not been included because observations cover
only out to $= R_{200}/2$. {\bf Right} EDisCS clusters (filled dots)
and literature data (empty dots, see text) over a radius $=
R_{200}/2$.
 \label{sfrmass}}
 \end{figure*}

The cluster-integrated SFR, normalized by the cluster mass (SFR/M) is
shown as a function of cluster mass in Fig.~\ref{sfrmass}.  The
cluster mass has been obtained from the cluster velocity dispersion
using eqn.~4 in Poggianti et al. (2006). Error bars are computed by
propagating the errors on the observed velocity
dispersion and the typical 10\% error on the [OII] flux. We reiterate
that these SFR estimates are not corrected for extinction.

From the Millennium Simulation, we find that mass and radius estimates
based on observed velocity dispersions critically fail for systems
below $\sim 300 \, \rm km \, s^{-1}$, yielding masses that are up to a
factor of 10 lower than the true virial mass of the system (Poggianti
et al. in prep.). As a consequence, the masses and the
mass-normalized SFRs for the two lowest velocity dispersion systems in
our sample (CL1119 and Cl1420) are likely to be blatantly incorrect,
and will not be used in the analysis. Nevertheless, for completeness
we do show the Cl1119 point in the diagrams. Cl1420 has no
galaxies showing [OII] emission.  It therefore has SFR=0 and is not
visible in the plots.

All of our other structures have SFR/M between 5 and 50 M$_{\odot}$
yr$^{-1}$ per $h^{-1}$ $10^{14}$ M$_{\odot}$.  Having excluded Cl1119
and Cl1420, the Kendall test gives a 95.7\% probability for an
anti-correlation between SFR/M and cluster mass (Fig.~\ref{sfrmass}).
Again without Cl1119 and Cl1420, the average SFR/M is 30.4 and 12.4
$M_{\odot}$ yr$^{-1}$ per $h^{-1}$ $10^{14}$ M$_{\odot}$ for systems below
and above $2\times 10^{14} \, h^{-1} \, {\rm M}_{\odot}$, respectively.

At redshift $\ge 0.4$ there are very few other clusters in the
literature with which we can compare. Cluster-integrated SFRs
corrected for incompleteness, within a clustercentric distance $=
R_{200}/2$, are presented by Finn et al. (2005) based on $\rm H\alpha$
studies for two additional clusters, Cl0024\footnote{For this cluster
we use the velocity dispersion given by Girardi \& Mezzetti (2001) to
derive the cluster mass.}  at $z=0.4$ (Kodama et al. 2004) and CL
J0023 at $z=0.85$.  A similar analysis was carried out by Homeier et
al.~(2005) for a cluster at $z=0.84$, except that it was based on
[OII] fluxes.  Both of these works, when including lower redshift
clusters, find a possible anti-correlation between the mass-normalized
cluster SFR and the cluster mass similar to ours, although it is
impossible to separate the redshift dependence from the mass
dependence in such small samples. An overall evolution of the
mass-normalized SFR and a large cluster-to-cluster scatter are also
found by Geach et al. (2006) using mid- to far-infrared data.
An upper limit in the mass-normalized SFR versus mass plane has
been found to exist for clusters, groups and individual galaxies by Feulner, 
Hopp \& Botzler (2006).

The right-hand panel of Fig.~\ref{sfrmass} shows SFR/M versus M for
the three clusters from the literature that were the subject of
emission-line studies, plotted alongside the EDisCS points restricted
to the same radius (=R$_{200} / 2$).  The SFRs for the non-EDisCS
clusters were corrected to account either for slightly different
SFR-[OII] calibrations or for the extinction of [OII] relative to $\rm
H\alpha$ (a factor of 2.5).
Including the three clusters from the literature and excluding Cl1119
and Cl1420 as above, we find that the average SFR/M is 33.4 and 9.8
$M_{\odot}$ yr$^{-1}$ per $h^{-1} \, 10^{14} \, M_{\odot}$ for systems
below and above $2\times 10^{14} \, h^{-1} \, M_{\odot}$,
respectively.  The Kendall test yields an anti-correlation probability
of 99.2\%.

\begin{figure*}[t]
\vspace{-6cm}
\centerline{\hspace{1cm}\includegraphics[width=0.6\columnwidth]{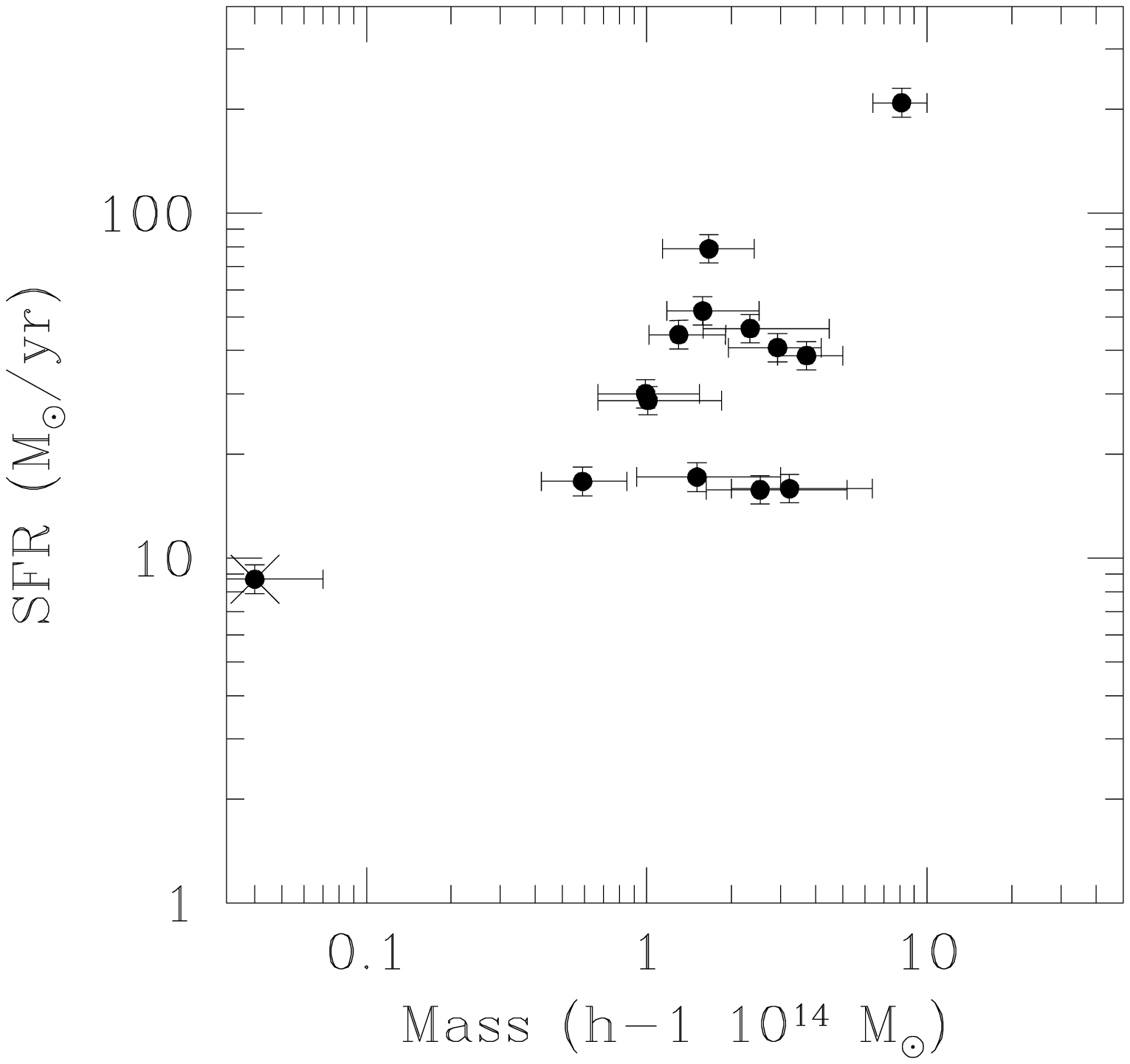}\hfill\hspace{-2cm}\includegraphics[width=0.6\columnwidth]{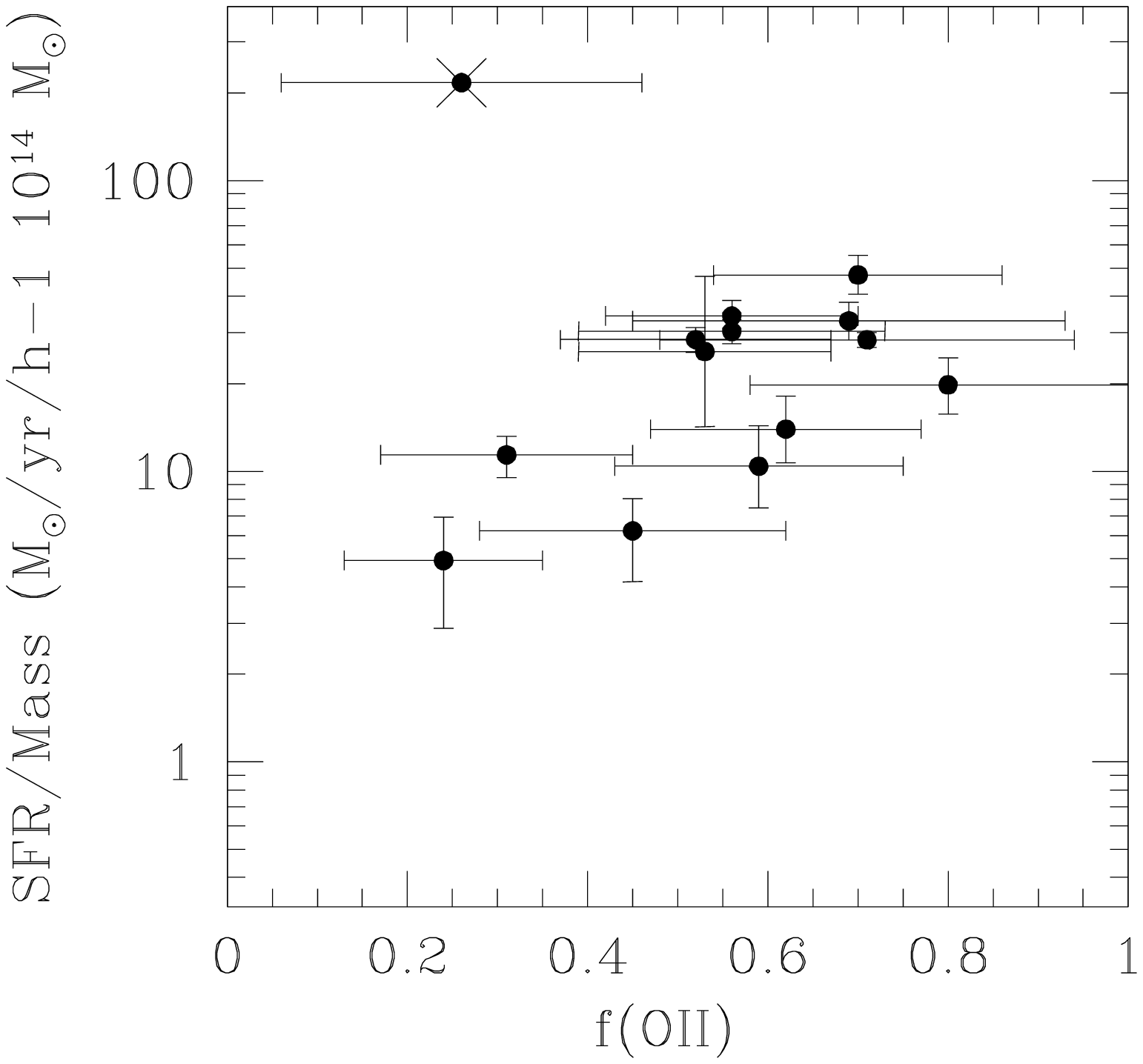}\hfill}
\caption{Total
integrated SFR as a function of cluster mass ({\bf right}) and
SFR/Mass versus fraction of star-forming
galaxies ({\bf left}). 
All quantities are computed within $R_{200}$. The large
cross identifies our lowest velocity
dispersion group (Cl1119) whose SFR/Mass estimate is
unreliable (see text).  Errors are computed by propagating errors on
velocity dispersions and SFRs.
 \label{sfroii}}
 \end{figure*}

In contrast, as shown in the left panel of Fig.~\ref{sfroii}, the
cluster-integrated SFR does not correlate with cluster mass (60\%
probability), and there is a large scatter in the mass range occupied
by the majority of our clusters ($1-5 \times 10^{14}$ $h^{-1} \, {\rm
M}_{\odot}$).  Moreover, the right panel of Fig.~\ref{sfroii} shows
that the SFR per unit mass follows the star-forming fraction (98\%).

We caution that the anticorrelation between SFR/M and M
presented in Fig.~\ref{sfrmass} could be entirely due to the
correlation of errors. We tested this possibility by generating 100
realizations of the dataset used in Fig.~\ref{sfroii} (i.e. 100
mass-SFR pairs), drawn from Gaussians with the same means and
intrinsic rms, and by adding Gaussian errors as observed. In 41 out of
100 cases the Kendall test gave a probability larger than 95.7\% that
an anticorrelation between mass and SFR/M exists. Therefore the
observed anticorrelation could be mainly driven by correlated errors,
although this test cannot rule out the existence of an intrinsic
anticorrelation.

Although our sample increases the number of available
cluster-integrated SFRs by a factor of four, larger cluster samples,
in particular clusters at the highest and lowest masses, are clearly
needed to verify our three findings: the weak anticorrelation of SFR/M
with M, the lack of a correlation between the integrated SFR and M,
and the presence of a correlation between SFR/M and star-forming
fraction.

\begin{figure*}[t]
\vspace{-1cm}
\centerline{\hspace{1.5cm}\includegraphics[width=0.4\columnwidth]{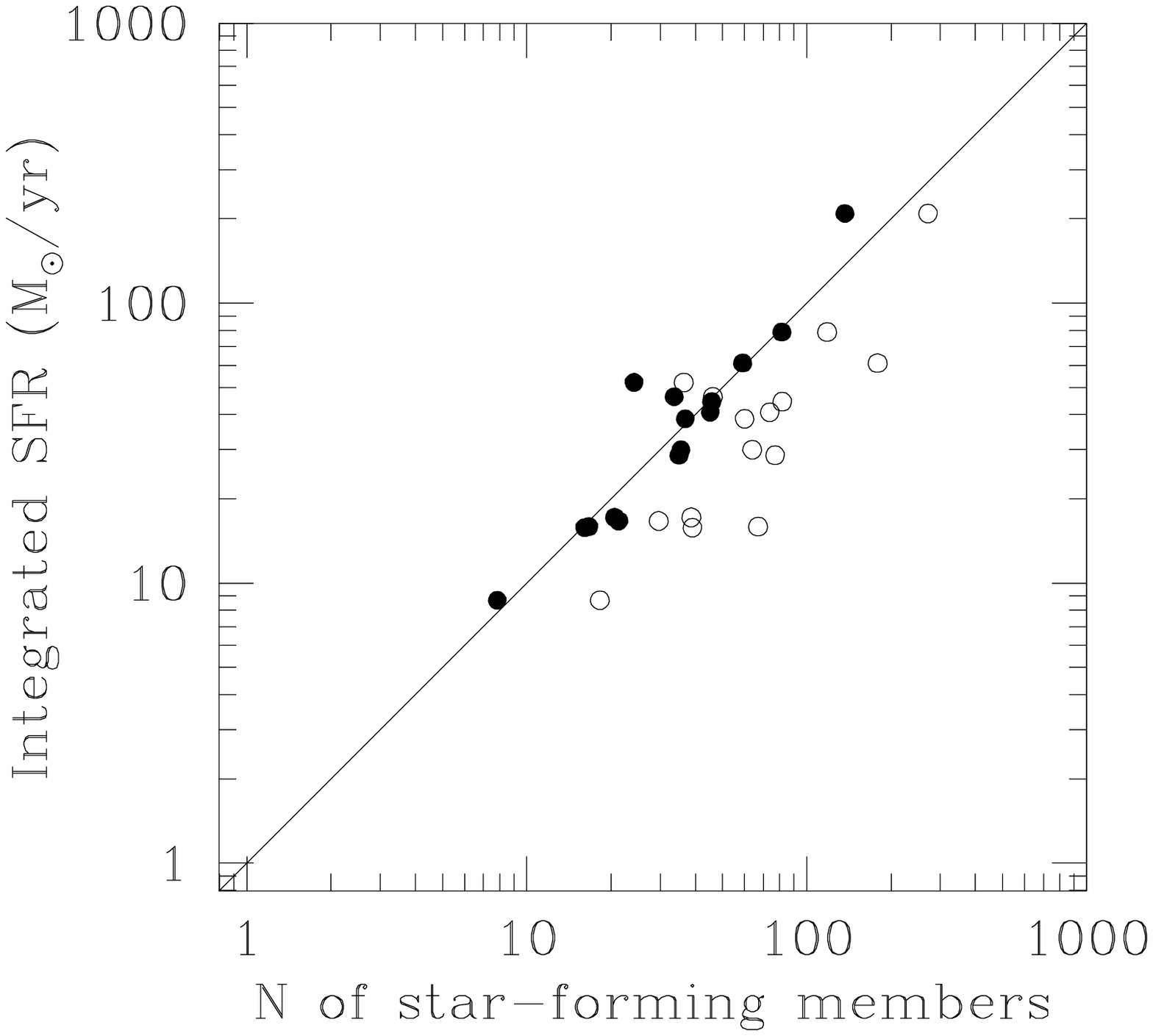}\hfill\hspace{-1.2cm}\includegraphics[width=0.4\columnwidth]{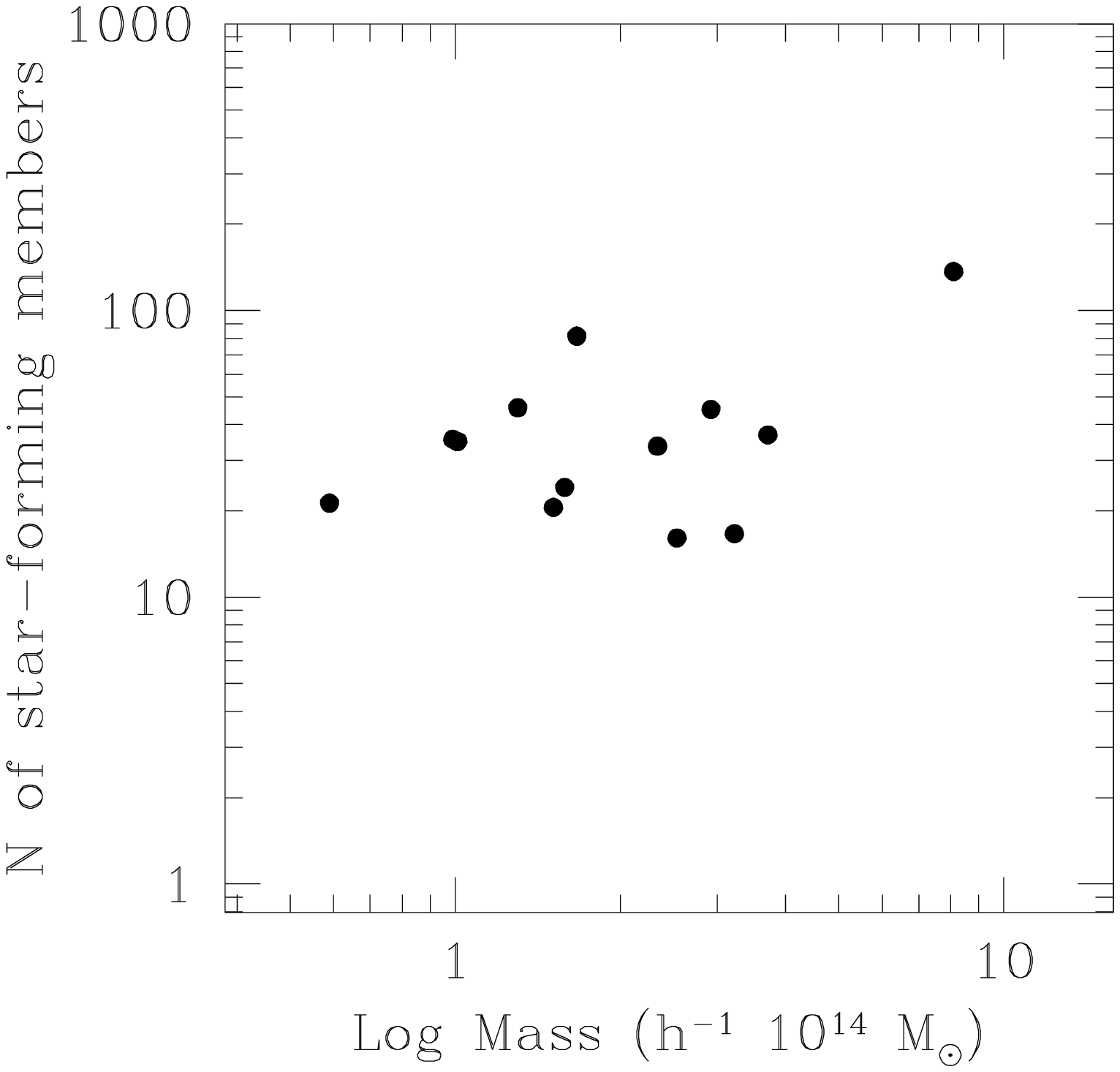}\hfill\hspace{-1.8cm}\includegraphics[width=0.4\columnwidth]{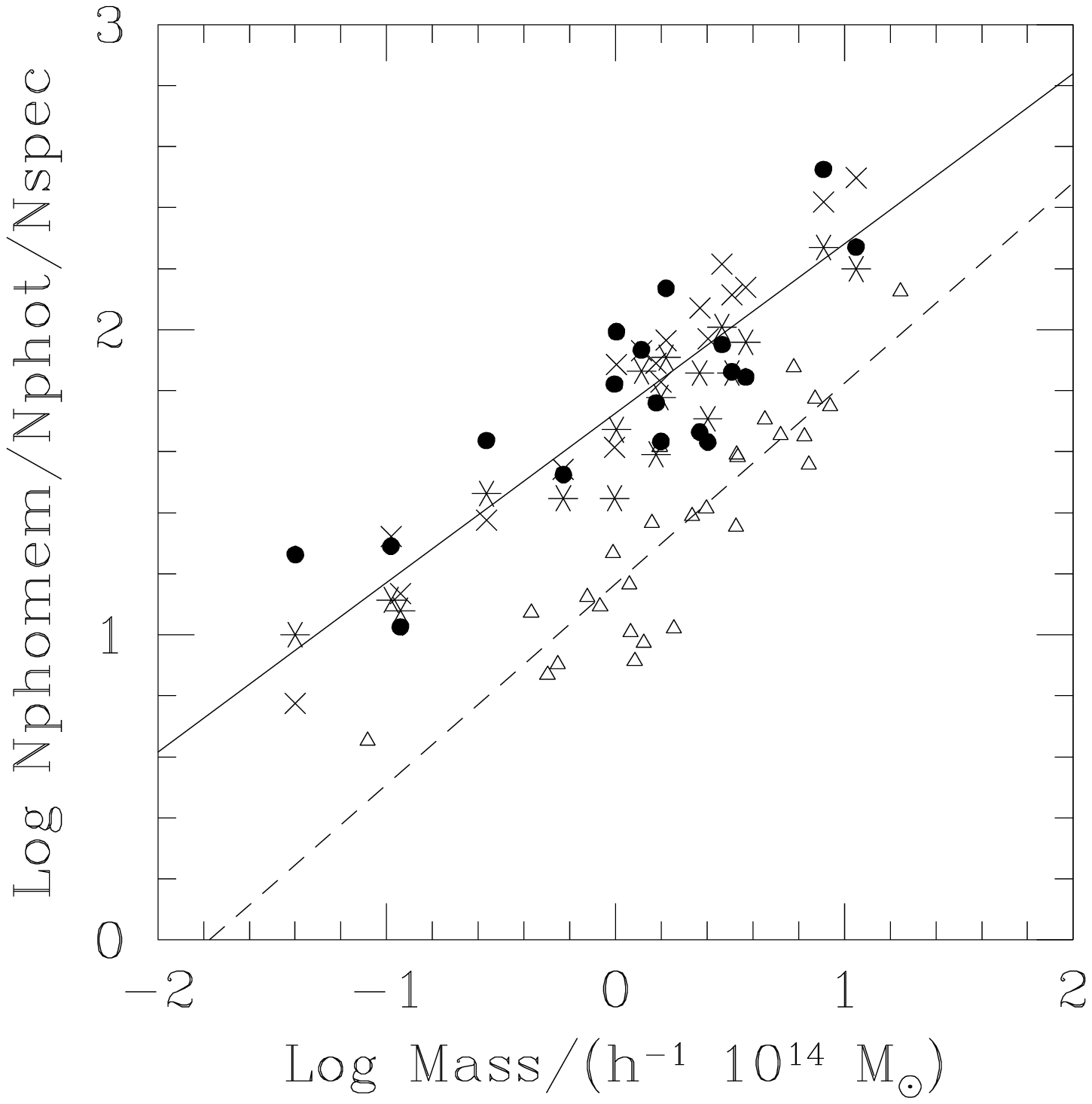}\hfill}
\caption{{\bf Left} Integrated cluster SFR versus number of star-forming
cluster members (filled circles) and total number of cluster members
(empty circles). The line represents the 1:1 relation,
$SFR = N_{SF}$, which implies that the average SFR per 
star-forming 
galaxy is roughly equal to $1 \rm \, M_{\odot} \, yr^{-1}$ in all clusters.
Numbers are computed using spectroscopically-confirmed members and correcting for spectroscopic incompleteness.
Error bars are omitted in this panel for clarity. All
quantities are computed within $R_{200}$ to the galaxy magnitude
limits adopted in this paper.  
{\bf Center} Number of star-forming
members versus cluster mass.
{\bf Right} Number of members within
$R_{200}$ versus cluster mass. For EDisCS clusters,
numbers are
computed from photo-z membership (stars), statistical substraction
(crosses) and from the number of spectroscopic members corrected for
incompleteness (filled circles).   
The least squares fit for EDisCS clusters is shown as a solid line,
and it is given in eqn.~1.
Sloan clusters at low-z are shown as empty triangles, and the least squares fit
as a dashed line (eqn.~3).
\label{sfrnum}}
 \end{figure*}

To further investigate the robustness and the possible origin of these
three results, in the left panel of Fig.~\ref{sfrnum} we show that the
integrated star formation is linearly proportional to the number of
star-forming galaxies $N_{\rm SF}$. In fact, the integrated SFR is
{\it equal} to the number of star-forming galaxies, because the
average SFR per star-forming galaxy is roughly constant in all
clusters at about $1 \rm \, M_{\odot} \, yr^{-1}$.  The correlation
between the integrated SFR and the number of star-forming galaxies is
much tighter than the relation between the SFR and the total number of
cluster members, also shown in Fig.~\ref{sfrnum} as empty circles.

In \pog we discovered that the star-forming fraction in distant
clusters generally follows an anticorrelation with cluster mass, with
some noticeable outliers, while in nearby clusters the average
star-forming fraction is constant for $\sigma > 500 \, \rm km \,
s^{-1}$, and increases towards lower masses with a large
cluster-to-cluster scatter.  The star-forming fraction is given by $f_{{\rm [OII]}}= N_{\rm SF}/N_{\rm tot}$.  In Fig.~\ref{sfrnum} we examine the mass dependence of both the numerator and denominator of this expression.  We show that 
in distant clusters the number of star-forming galaxies $N_{\rm SF}$
does not depend on cluster mass (central panel), while the total
number of cluster members $N_{\rm tot}$ grows with cluster mass (right
panel) according to a least squares fit as:

\begin{equation}
log(N_{tot}) = 0.56 \times log M(h^{-1} \, 10^{14} \, M_{\odot}) +
1.73
\end{equation} 

At $z=0$, the star-forming fraction in systems more massive than $500
\, \rm km \, s^{-1}$ is constant.  If the average star formation
activity {\it in star-forming galaxies} in these clusters is
independent of cluster mass, as it is at high redshift, then the
cluster-integrated star formation rate at $z=0$ should be not only
linearly proportional to the number of star-forming galaxies, but also
to the total number of cluster members, as indeed found by Finn et
al.~(2008).

Moreover, in low-z clusters the relation between the total number of
cluster members and cluster mass is (triangles in the right panel of
Fig.~\ref{sfrnum})

\begin{equation}
\log(N_{\rm tot}) = 0.66 \times \log M(h^{-1} \, 10^{14} \, M_{\odot}) + 1.17
\end{equation} 

As a consequence of eqn.~(2) and of the constancy of the star-forming
fraction presented in \pog, the number of star-forming galaxies in
clusters with $\sigma > 500$ $\rm km \, s^{-1}$ at low-z, as well as
the total integrated star formation rate, must increase with cluster
mass. This is at odds with what we find at high-z, where both the
number of star-forming galaxies and the total SFR are independent of
cluster mass (Fig.~\ref{sfrnum} and Fig.~\ref{sfroii}).  The different
behavior at $z=0.6$ and $z=0$ is simply due to the different trends
of the star-forming fraction with cluster mass at the two redshifts
(\pog).

In systems with masses below $500 \, \rm km \, s^{-1}$ at $z=0$,
$N_{]rm SF}/N_{\rm tot}$ is no longer independent of cluster mass,
being on average (\pog)

\begin{equation}
N_{\rm SF}/N_{\rm tot} = -2.2 \times \frac{\sigma}{1000 \, \rm km \, s^{-1}} + 1.408 
\end{equation}

Based on eqns.~2 and 3 and the relation between cluster
mass and $\sigma$, we predict that 
the average number of star-forming galaxies for low-mass systems at
$z=0$
should be equal to between 4 and 6 galaxies regardless of group mass
for masses between $2 \times 10^{13}$ and $2 \times 10^{14} \, h^{-1}
\, M_{\odot}$.  {\it If} the average star formation rate per
star-forming galaxy 
is independent of group mass 
at low-z, as it is at high-z, then 
the
{\it average} total group star formation rate in the mass range $2 \times 10^{13}-2
\times 10^{14} \, h^{-1} \, M_{\odot}$ 
should also be constant,
with a very large scatter from group to group at a given mass
reflecting the large scatter in the star-forming fraction. Low-z group
samples should be able to verify these predictions, which are based
purely on the observed correlations presented in this paper and in
\pog.

Because the integrated SFR is equal to the number of star-forming
galaxies in distant clusters,
the former is by definition proportional (with a proportionality
factor that happens to be equal to 1) to the star-forming fraction
multiplied by the total number of cluster members ${\rm SFR} = N_{\rm
SF} = f(OII) \times N_{\rm tot}$.  In distant clusters, the best-fit
relation between $f(OII)$ and cluster mass was given by \pog:

\begin{equation}
f(OII)=N_{\rm SF}/N_{\rm tot} = -0.74 \times \frac{\sigma}{1000 \, \rm km \, s^{-1}} + 1.115 
\end{equation}
  
From this and from the fact that
the total number of cluster
galaxies correlate with cluster mass (eqns. 4 and 1), and given
that the integrated SFR is equal to the number of star-forming
galaxies (Fig.~\ref{sfrnum}), one can analytically
 conclude that the SFR/M should
correlate with the star-forming fraction, as indeed we observe in
Fig.~\ref{sfroii}.

To summarize, in distant clusters we have observed a weak
anticorrelation between SFR/M and cluster mass, the lack of any
correlation between cluster-integrated SFR and mass, and the presence
of a correlation between SFR/M and star-forming fraction.  These
findings can be explained, and actually predicted, on the basis of
three observed quantities: a) the constancy of the average SFR per
star-forming galaxy in all clusters, found in this paper (left panel
of Fig.~\ref{sfrnum}); b) the correlation between cluster mass and
number of member galaxies shown in this paper (eqn.~1 and right panel
of Fig.~\ref{sfrnum}); and c) the previously observed dependence of
star-forming fraction on cluster mass (\pog).

Observation (a), that the average SFR per star-forming galaxy is
constant for clusters of all masses, suggests that either clusters of
all masses affect the star formation activity in infalling
star-forming galaxies in the same way, or that, if/when 
they cause a truncation of the star formation, they do so
on a very short timescale.  
In the latter case, star-forming galaxies of a given
mass have similar properties inside and outside of clusters.
Observation (b), the correlation between cluster mass and number of
cluster members, stems from the mass and galaxy accretion history of
clusters. 
These results can be used to test the predictions of simulations.
More importantly, comparisons with simulations can allow to explore how
our results are linked with the growth history of clusters, which
should play an important role in establishing the star-forming
fraction.
The relative numbers of star-forming versus non-star-forming galaxies
(observation (c) above) and, above all, its evolution, remain the key
observations that display a strong dependence on cluster
mass. Ultimately, understanding the observed trends comes down to
finding out why the relative proportion of passive and star-forming
galaxies varies with ``environment', the latter being either cluster
mass or local density. In \pog we proposed a schematic scenario in
which there are two channels that cause a galaxy to be passive in
clusters today: one due to the mass of the galaxy host halo at $z>2$
(a ``primordial'' effect), and one due to the effects related to the
infall into a massive structure (a ``quenching'' mechanism). The
results of this paper are consistent with that simple picture.

\section{Age or morphology?}

For 10 EDisCS fields we can study galaxy morphologies from visual
classifications of HST/ACS images (Desai et al. 2007) and thus compare
our star formation estimates with galaxy Hubble types. In particular,
we are interested in knowing whether the trend of SF with local
density can be partially or fully ascribed to the existence of a
morphology-density relation (MDR).  Do the SF trends simply reflect a
different morphological mix at different densities, with the SF
properties of each Hubble type being invariant with local density? Or
do the SF properties of a given morphological type depend on density?
Can the lower average SF activity in denser regions be fully explained
by the higher proportion of early-type galaxies in denser regions?

Desai et al.~(2007) have published visual classifications in the form
of Hubble types (E, S0, Sa, Sb, Sc, Sd, Sm, Irr).  However, for this
paper we consider only four broad morphological classes: E
(ellipticals), S0s (lenticulars), early-spirals (Sa's and Sb's) and
late-spirals (Sc's and later types).  We note that irregular galaxies
(Irr) represent only 10\% of our late-spiral class and therefore do
not dominate any of the late-spiral results we present below.

\begin{figure*}[t]
\centerline{\includegraphics[width=13cm]{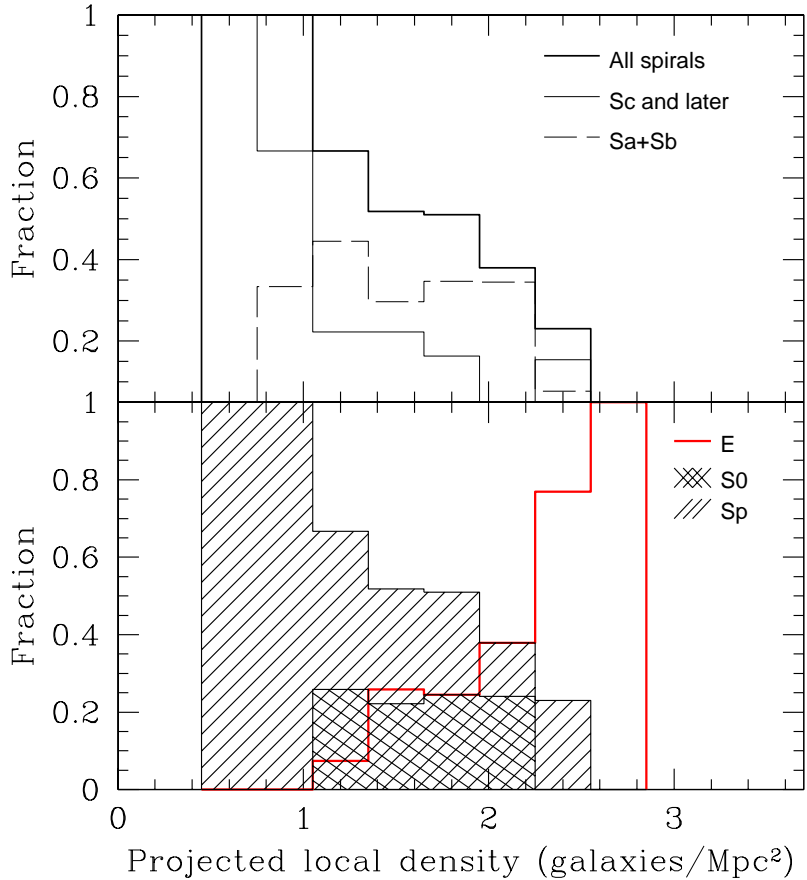}}
\caption{Morphology-density relation for EDisCS
spectroscopically-confirmed cluster
members. Fraction of each morphological
type in various density bins. {\bf Bottom} Es: thick black
histogram. S0: double shaded histogram. Spirals: single shaded
histogram. {\bf Top} All spirals: thick solid histogram. Early-type
spirals (Sa's and Sb's): long-dashed histogram. Late-type spirals
(Sc's and later types): thin solid histogram.  Note that the highest
density bin is only populated by ellipticals.
 \label{md}}
 \end{figure*}

The morphology-density relation for EDisCS spectroscopically-confirmed
cluster members brighter than $M_V=-20$ is shown in Fig.~\ref{md}. We
find clear trends similar to what has been observed before in clusters
both at high- and low-z (Dressler et al. 1997, Postman et al. 2005,
Smith et al. 2005).  The fraction of spirals decreases, the fraction
of ellipticals increases, and the fraction of lenticulars is flat with
local density.

Previous high-z studies have not considered early-spirals and
late-spirals separately. We find that the spiral trend is due to the
fraction of late-spirals strongly decreasing with density, while the
distribution of early-spirals is rather flat with density (top panel
of Fig.~\ref{md}). The early-spiral density distribution is thus very
similar to that of S0 galaxies, suggesting that these objects are the
best candidates for the immediate progenitors of the S0 population,
which has been observed to grow between $z=0.5$ and $z=0$ (Dressler et
al. 1997, Fasano et al. 2000, Postman et al. 2005, Desai et al. 2007).

We consider three observables related to the star formation activity:
the [OII] EW and the SFR derived from the [OII] flux described in the
previous sections, and the break at 4000 \AA.  The last is defined as
the difference in the level of the continuum just bluer and just
redder than 4000 \AA.  It can be thought of as a ``color'' and in fact
it usually correlates well with broad band optical colors, though it
spans a smaller wavelength range than broad bands and is thus less
sensitive to the dust obscuring those stars that dominate the spectrum
at these wavelengths. We use the narrow version of this index, sometimes
known as $D4000n$, as defined by Balogh et al. (1999), and refer to it
as D4000 in the following.

\begin{figure*}[t]
\centerline{\hspace{1cm}\includegraphics[width=0.7\columnwidth]{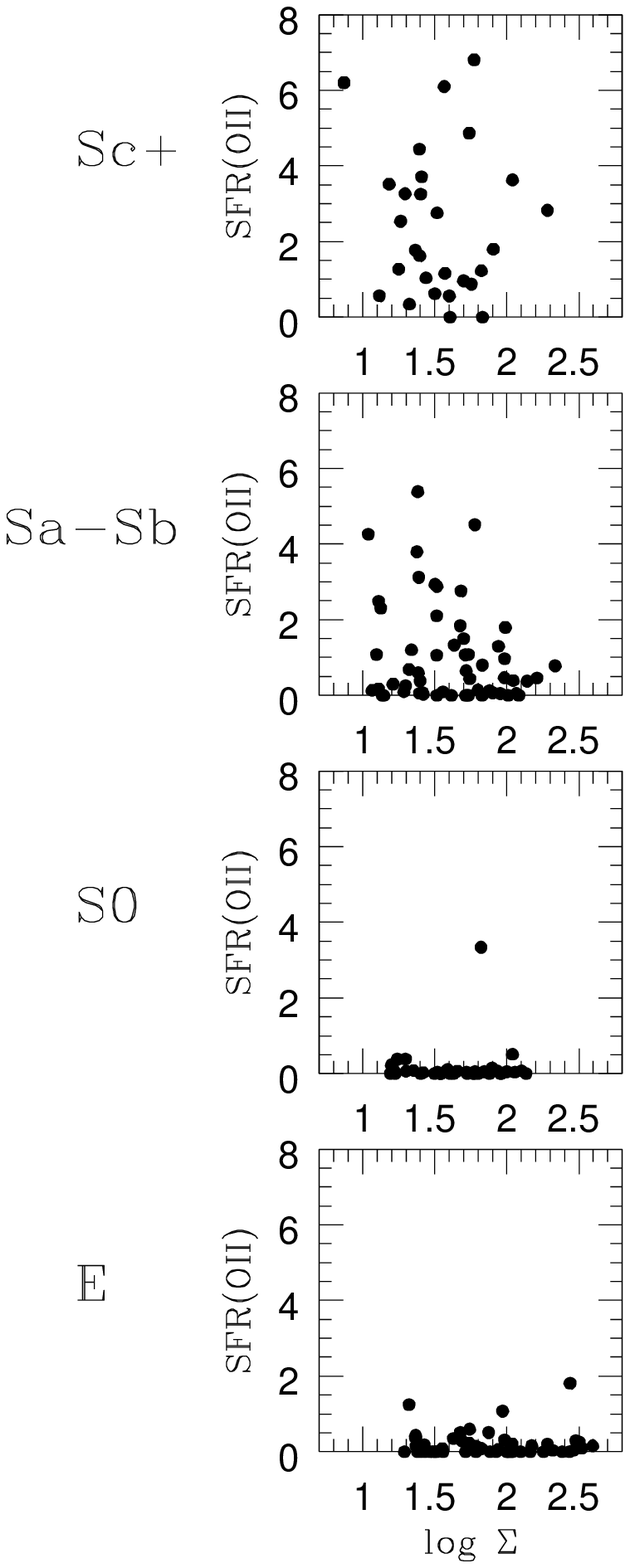}\hfill\hspace{-8cm}\includegraphics[width=0.7\columnwidth]{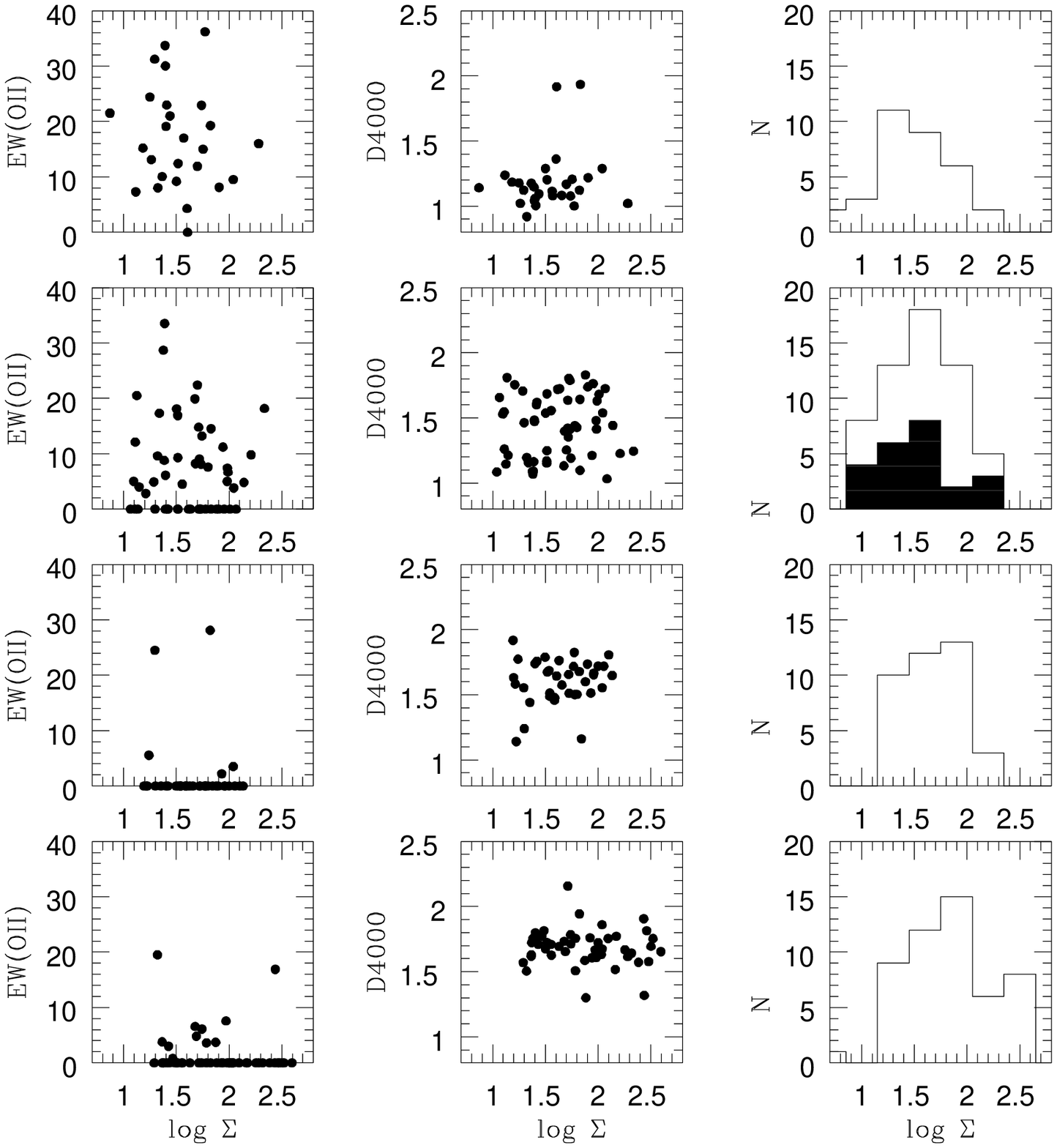}}
 \caption{``Age-density'' relation for galaxies of different
 morphological types. From left to right: SFR, [OII] equivalent width,
 D4000, and number of galaxies versus local density for (from bottom to
 top) Es, S0s, early spirals, and late spirals. The filled histogram
 shows the density distribution of ``blue'' early spirals with
 $D4000<1.3$.
 \label{ageden}}
 \end{figure*}

The observed values of SFR, EW(OII), and D4000 are plotted as a
function of local density for each of our four morphological classes in
Fig.~\ref{ageden} (Es, S0s, early-spirals, late-spirals from bottom to
upper row). The same figure presents the local density distribution of
each morphological class. The highest density bin is only populated by
ellipticals, as previously discussed.

Two main conclusions can be drawn from this figure: 

a) Neither the SFR, nor the EW(OII), nor the D4000 distributions of each
given morphological class vary systematically with local density.
Dividing each morphological class into two equally-populated density
bins, we find statistically consistent mean and median values of SFR,
EW([OII]), and D4000.  The only exception may be a possible deficiency
of galaxies with high SFR among early spirals at the highest
densities.  However, the mean and median SFR in the two density bins
differ only at the 1$\sigma$ level.  Thus, as far as it can be
measured in our relatively small sample of galaxies, the SF properties
of a given morphological class do not depend on density.

b) While the great majority of Es and S0s are ``red'' (==have high
values of D4000, and null values of EW(OII) and SFR) and the great
majority of late-type spirals are ``blue'' (==have low D4000 values,
OII in emission and ongoing SF), early-spirals are a clearly bimodal
population composed of a red subgroup ($D4000>1.5$) and a blue
subgroup ($D4000<1.3$). Approximately 40\% of the early-spirals are
red with absorption-line spectra and 40\% are blue with emission-line
spectra, and the rest have intermediate colors.

Most of the intermediate-color galaxies have some emission, but at
least two out of 10 have recently stopped forming stars (have k+a
post-starburst spectra; Poggianti et al.~2008 in prep.) and therefore
are observed in the transition phase while moving from the blue to the
red group.
For their star-formation properties, the red early-spirals can be assimilated
 to the
``passive spirals'' observed in several previous surveys
(Poggianti et al. 1999, Moran et al. 2006, Goto et al. 2003,
Moran et al. 2007).
 
The early-spiral bimodality is not due to the red subgroup being
composed mainly of Sa's and the blue subgroup consisting mainly of
Sb's, as the proportion of Sa's and Sb's is similar in the two
subgroups.  Interestingly, the relative fractions of ``red'' and
``blue'' early-spirals does not strongly depend on density, as might
have been expected, but there is a tendency for the intermediate color
galaxies to be in regions of high projected local density.

We now want to calculate whether the observed star formation-density relations
can be accounted for by the observed morphology-density relation,
combined with the average SF properties of each morphological class.

\begin{figure}[t]
\vspace{-8cm}
\centerline{\hspace{2cm}\includegraphics[width=0.8\columnwidth]{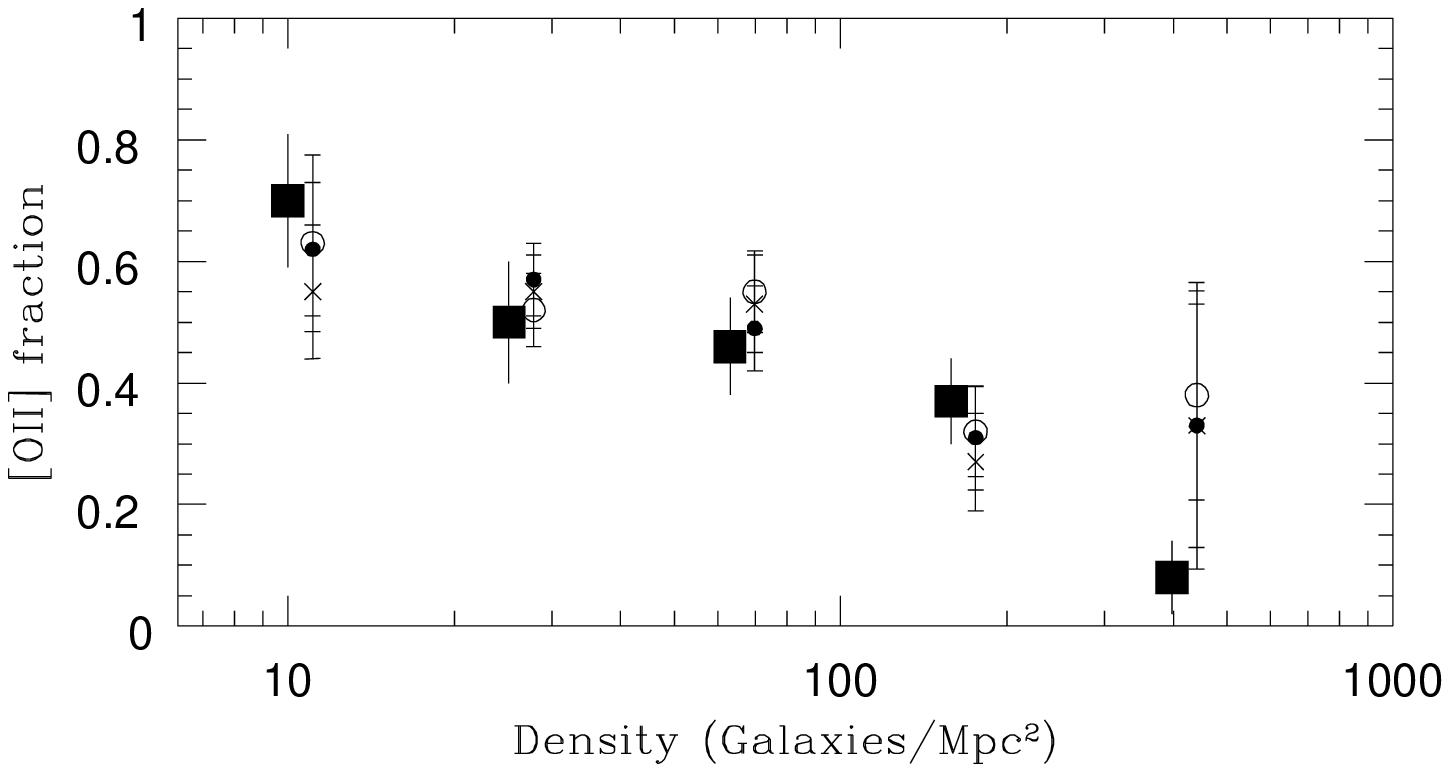}}
\vspace{-8cm}
\centerline{\hspace{2cm}\includegraphics[width=0.8\columnwidth]{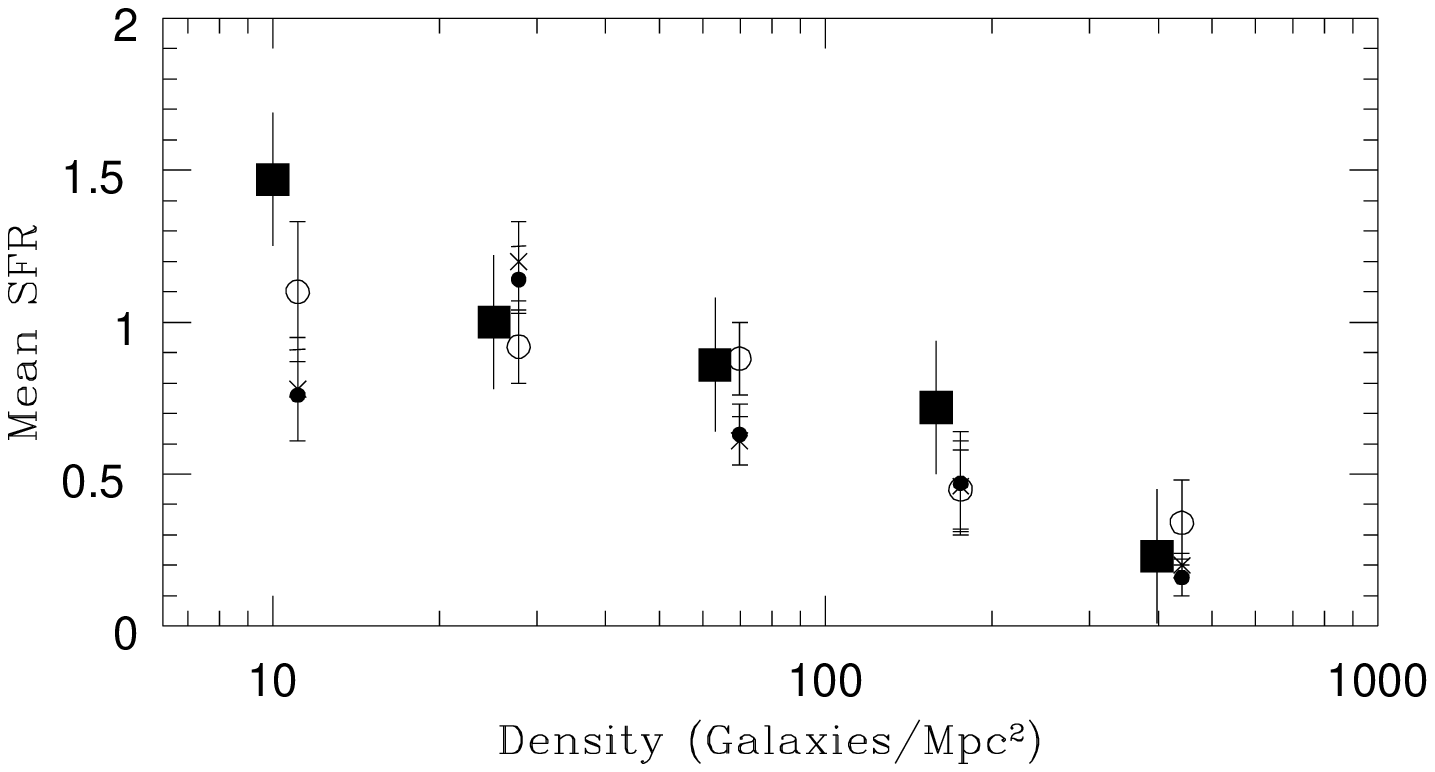}}
\vspace{-1cm}
 \caption{Fraction of star-forming galaxies (top) and mean SFR among all
galaxies (bottom) with symbols as in Fig.~\ref{fractions} and
Fig.~\ref{mainsfr}. The large, solid squares represent the values
expected given the morphology-density relation and the mean
star-forming fraction and SFR of galaxies of each morphological class.
 \label{mainmorph}}
 \end{figure}

To obtain the trend of star-forming fraction with density expected
from the MDR, we compute the fraction of star-forming galaxies in each
morphological class and combine this with the fraction of each
morphological class in each density bin (i.e. the MDR).\footnote{Note
that Hubble types are known only for a subset of our clusters, thus
the sample used in this section and shown as large symbols is a
subsample of the whole spectroscopic sample used for the total SF
relations, yet the latter are well reproduced.}  The result is
compared with the observed star-forming fractions in
Fig.~\ref{mainmorph}.

Similarly, to compute the expected SFR-density relation given the
MD-relation, we combine the mean SFR in solar masses per year for each
morphological class (0.23$\pm$0.1 for Es, 0.15$\pm$0.1 for S0s,
1.03$\pm$0.16 for early-spirals and 2.71$\pm$0.43 for late spirals)
with the fraction of each morphological class in each density bin
(i.e. the MDR), and compare it with the observed SFR-density relation
in Fig. \ref{mainmorph}.

This figure shows that the MDR is able to fully account for the
observed trends of star-forming fraction and SFR with density (and
viceversa).\footnote{Note that accounting for the star-forming
fraction as a function of density means also accounting for the mean
EW(OII) trend with density for all galaxies, given the constancy of
mean EW(OII) with density for star-forming galaxies.} Hence, we find
that the MD relation and the ``star formation-density relation'' (in
the different ways it can be observed) are equivalent. These
observations indicate that at least in clusters, for the densities,
redshifts, galaxy magnitudes, SF and morphology indicators probed in
this study, these are simply two independent ways of observing the
same phenomenon, and that neither of the two relations is more
``fundamental'' than the other.

\subsection{Comparison with low redshift results}

The equivalence between the SFD and the MD relations that we find in
EDisCS clusters is at odds with a number of studies at low redshift.
In local clusters, Christlein \& Zabludoff (2005) have found a
residual correlation of current star formation with environment
(clustercentric distance in their case) for galaxies with comparable
morphologies and stellar masses. Using the Las Campanas Redshift
Survey, Hashimoto et al.~(1998) demonstrated that the star formation
rates of galaxies of a given structure depend on local density, and
that ``the correlation between...star formation and the bulge-to-disk
ratio varies with environment''.  More recently, a series of works on
other field low-z redshift surveys have concluded that the star
formation-density relation is the strongest correlation of galaxy
properties with local density, suggesting that the most fundamental
relation with environment is the one with star formation histories,
not with galaxy structure (Kauffmann et al. 2004, Blanton et al. 2005,
Wolf et al. 2007, Ball, Loveday \& Brunner 2008).  Most of these
studies are based on structural parameters such as concentration or
bulge-to-disk decomposition that are used as a proxy of galaxy
``morphology''. Visual morphologies such as those we use in this paper
are known to be related both to structural parameters and star
formation. In fact, from a field SDSS galaxy sample at low-z, van der
Wel (2008) concludes that structure mainly depends on galaxy mass and
morphology depends primarily on environment, and that the MD relation
at low-z is ``intrinsic and not just due to a combination of more
fundamental, underlying relations''. Similarly, Park et al. (2007)
argue that the strongest dependence on local density is that of
morphology, when morphology is defined by a combination of
concentration index, color and color gradients. Interestingly, having
fixed morphology and luminosity, these authors find that both
concentration {\it and} star formation related observables are nearly
independent of local density.

\begin{figure}[t]
\vspace{-5cm}
\centerline{\hspace{1cm}\includegraphics[width=12cm]{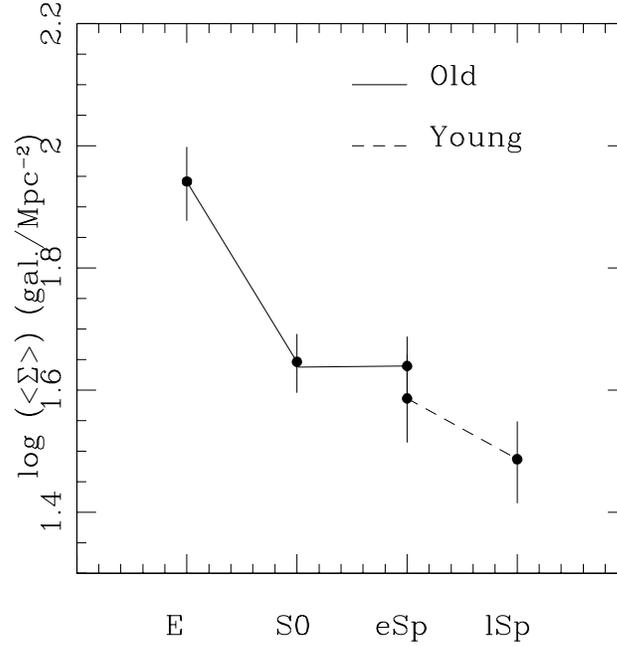}}
\vspace{-1cm}
 \caption{Mean local density for each morphological type, from left to
right: ellipticals, S0s, early spirals, and late spirals.  The notation
``old'' and ``young'' here separates galaxies with red and blue D4000 ($>/<
1.3$). Practically all ellipticals and S0s are old, and all
late spirals are young, while early spirals are cleanly divided into two
age populations with similar mean densities. Errors are bootstrap standard 
deviations.
 \label{wolf}}
 \end{figure}

The difference between a classification based on structural parameters
and one obtained from visual morphology may be responsible for the
differences between our high-z results and most, but not all,
low-z results. Using visual morphologies of galaxies in the
supercluster A901/2, Wolf et al.~(2007) find that the mean projected
density of galaxies of a given age does not depend on morphological
class, and conclude there is no evidence for a morphology-density
relation at fixed age. In their sample, except for the latest spirals,
which are all young, galaxies of every other morphological type span
the whole range of ages, i.e. there are old, intermediate-age, and
young Es, S0s, and early spirals.  In contrast, as discussed
previously, our morphological classes correspond to a strong
segregation in age: practically all ellipticals and S0s are old, all
late spirals are young, and only early spirals are a bimodal
population in age. To facilitate the comparison with Wolf et
al. (2007), in particular with their Fig.~5c, in Fig.~\ref{wolf} we
present our results as mean projected density for galaxies of
different stellar ``ages'' as a function of morphological class. The
notation ``old'' and ``young'' separates galaxies with red and blue
D4000 ($>/< 1.3$).  

Figure \ref{wolf} shows that 
we find an ``MD-relation'' at fixed
age, i.e. a difference in mean density for galaxies of the same age
but different morphological type, for example between old Es and old
S0s.

Since age trends with density are observed only at faint magnitudes by
Wolf et al., the fact that their galaxy magnitude limit is 2 mag deeper
than ours may partly or fully explain the discordant conclusions.
The SFD and the MD relations may be equivalent at bright magnitudes,
and decoupled at faint magnitudes.  

Additionally, it is possible that we are observing an evolutionary
effect, with star formation and morphology equally depending on
density at high-z, but not at low-z. This might be the case if at
high-z most galaxies still retain the morphological class they had
``imprinted'' in the very early stages of their formation, and if at
lower redshifts progressively larger number of galaxies are
transformed, having their star formation activity and morphology
changed.  Such transformations are known to have occurred in a
significant fraction of local cluster galaxies.  
In fact, approximately 60\% of
today's galaxies have evolved from star-forming at $z \sim 2$ to
passive at $z=0$ according to the results of \pog.

If the changes in star formation are more closely linked with the
local environment than the related change in morphology, while the
latter retains some memory of the initial structure at very high-z
(mostly dependent on galaxy mass), a progressive decoupling between
the SFD and the MD relations would take place at lower redshifts (see
also Capak et al. 2007).  Since the changes in star formation activity
and morphology involve progressively fainter galaxies at lower
redshifts in a downsizing fashion (Smail et al. 1998, Poggianti et
al., 2001a,b, 2004, De Lucia et al. 2004, 2007), the decoupling at
low-z should be prominent at faint magnitudes.

In this scenario, the differences between our analysis and Wolf's
results would be both an evolutionary and a galaxy magnitude limit
effect, the two being closely linked. At low-z it should now be
possible to fully address these questions and investigate the galaxy
magnitude and global environment dependence of the SFD-MD decoupling.
Ours is so far the only study comparing the SFD and the MD relations
at high redshift, so other future works may help clarify the redshift
evolution of the link between the two relations in clusters, groups,
and the field.

\section{Summary}

We have measured the dependence of star formation activity and
morphology on projected local galaxy number density for cluster,
group, poor group, and field galaxies at $z=0.4-0.8$, comparing with
clusters at low redshift. At high-z, our 16 main structures have
measured velocity dispersions between 160 and 1100 $\rm km \, s^{-1}$,
while for our poor groups we did not attempt a velocity dispersion
measurement. The field sample comprises galaxies that do not belong to
any of our clusters, groups, or poor groups.

Our analysis is based on the [OII] line equivalent widths and fluxes
and does not include any correction for dust extinction. All galaxies
with an EW([OII]) greater than 3 \AA $\,$ are considered to be currently
star-forming. Although the contamination from pure AGNs is estimated
to be modest (7\% at most), all the trends shown might reflect a 
combination of both star formation and AGN activity.

Our main conclusions are as follows:

1) In distant as in nearby clusters, regions of higher projected
density contain proportionally fewer galaxies with ongoing star
formation. Both at high and low redshift, the average star formation
activity in star-forming galaxies, when measured as mean [OII]
equivalent width, is consistent with being independent of local
density.

2) At odds with low-z results, we find that the correlation between
star-forming fraction and projected local density varies for massive
and less massive clusters, though it is not uniquely a function of
cluster mass. Some low-mass groups can have lower star-forming
fractions at any given density than similarly or more massive
clusters.

3) In our clusters, the average current star formation rate
per galaxy and per star-forming galaxy, as well as the average star
formation rate per unit of galaxy mass, do not follow a continuously
decreasing trend with density, and may display a peak at densities
$\sim 15-40$ galaxies Mpc$^{-2}$. The significance of this peak ranges
between 1 and 4 $\sigma$ depending on the method of analysis.  This
result could be related to the recent findings of an inverted,
possibly peaked SFR-density relation in the field at $z=1$.

The EW-density and the SFR-density relations thus provide different
views of the correlation between star formation activity and
environment. The former suggests that the star formation activity in
star-forming galaxies does not vary with local density, while the
latter suggests the existence of a density range in which the star
formation activity in star-forming galaxies is boosted by a factor
of $\sim$1.5 on average.

4) When using galaxy samples with similar mass distributions, we find
variations not larger than 1 $\sigma$ in the average EW and SFR
properties of star-forming galaxies in the field, poor groups, and
clusters.  Higher average EWs, SFRs, and star-forming fractions in the
unmatched field and poor group samples compared to clusters are
primarily due to differences in the galaxy mass distribution with
global environment. As an example, star-forming field galaxies form
stars at more than twice the rate per unit of galaxy mass compared to
star-forming galaxies in any other environment.  Together with point
1) above, this suggests that the current star formation activity in
star-forming galaxies of a given galaxy mass does not strongly depend
on global or local environment.


5) By summing the ongoing SFR of individual galaxies within each
cluster we obtain cluster-integrated star formation rates.  We find
no evidence for a correlation with cluster mass. In contrast, the
cluster SFR per unit of cluster mass anticorrelates with mass and
correlates with the star-forming fraction, although we caution
that the anticorrelation with mass could be mainly driven
by correlated errors. 
 The average star-forming
galaxy happens to form about one solar mass per year (uncorrected for
dust) in all of our clusters, making the integrated star formation
rate in distant clusters just equal to the number of star-forming
galaxies.

These findings can be understood in the light of three additional
results that we show: a) the cluster integrated SFR is linearly
proportional (equal) to the number of star-forming galaxies; b) the
total number of cluster members scales with cluster mass as $N \propto
M^{0.56}$ and c) the star-forming fraction depends on cluster mass in
distant clusters as presented in \pog.  Given the invariance of the
average star formation with cluster mass, as well as with global and
local environment (see points 1 and 4 above), the most important
thing that remains to be explained is the cause of the
cluster-mass-dependent evolution of the relative number of
star-forming versus non-star-forming galaxies.

6) Defining galaxy morphologies as visually classified Hubble types
from HST/ACS images, we find a morphology-density relation similar to
that observed in previous distant cluster studies. In addition, we
find that the trend of declining spiral fraction with density is
entirely driven by late-type spirals of types Sc and later, while
early spirals (Sa's and Sb's) have a flat distribution with local
density as S0s do.

7) The star formation properties (ongoing SFR, EW(OII), and D4000) of
each morphological class do not depend on local density.  Galaxies of
a given Hubble type in distant clusters have similar star formation
properties regardless of the local environment.

8) Essentially all Es and S0s have old stellar populations and all
late spirals have significant young stellar populations, while early
spirals are a clearly bimodal population, with 40\% of them being red
and passively evolving and 40\% being blue and having ongoing star
formation.  The bimodality of the early spirals, together with their
resemblance to S0s as far as the morphology-density distribution is
concerned, once more suggests that early spirals are the most
promising candidates for the progenitors of a significant fraction of
the S0 population in clusters today (see also Moran et al. 2007).

9) From the combination of the morphology-density relation and the
average properties of each morphological class, we are able to recover
the star formation-density relations we have observed. The
morphology-density and the star formation-density relation are
therefore equivalent in our distant clusters, and neither of the two
relations is more fundamental than the other.  This is at odds with
recent results at low-z. Among the possible reasons for the discordant
conclusions are differences between visual morphologies and structural
parameters, the fainter galaxy magnitude limit reached in low-z
studies, and possibly evolutionary effects that can produce a
progressive decoupling of the SFD and the MD relations at
lower redshifts.

\acknowledgments 
We would like to thank the referee, Arjen van der
Wel, for the constructive and careful report that helped us improving
the paper.  BMP thanks the Alexander von Humboldt Foundation and the
Max Planck Instituut fur Extraterrestrische Physik in Garching for a
very pleasant and productive stay during which the work presented in
this paper was carried out.  The Millennium Simulation databases used
in this paper and the web application providing online access to them
were constructed as part of the activities of the German Astrophysical
Virtual Observatory.  The Dark Cosmology Centre is funded by the
Danish National Research Foundation. BMP acknowledges financial
support from the FIRB scheme of the Italian Ministry of Education,
University and Research (RBAU018Y7E) and from the INAF-National
Institute for Astrophysics through its PRIN-INAF2006 scheme.



{\it Facilities:} \facility{VLT (FORS2)}, \facility{HST (ACS)}.

\end{document}